\documentclass[12pt,a4paper]{article}\usepackage[]{graphicx}\usepackage[]{color}
\makeatletter
\def\maxwidth{ %
  \ifdim\Gin@nat@width>\linewidth
    \linewidth
  \else
    \Gin@nat@width
  \fi
}
\makeatother

\definecolor{fgcolor}{rgb}{0.345, 0.345, 0.345}

\usepackage{framed}
\makeatletter
 {\par\unskip\endMakeFramed%
 \at@end@of@kframe}
\makeatother

\definecolor{shadecolor}{rgb}{.97, .97, .97}
\definecolor{messagecolor}{rgb}{0, 0, 0}
\definecolor{warningcolor}{rgb}{1, 0, 1}
\definecolor{errorcolor}{rgb}{1, 0, 0}
\newenvironment{knitrout}{}{} % an empty environment to be redefined in TeX

\usepackage{alltt}
\usepackage[utf8]{inputenc} %PDFLaTeX
\usepackage[T1]{fontenc} %PDFLaTeX
\usepackage{csquotes}
\usepackage{amsmath}
\usepackage{booktabs}
\usepackage[font=scriptsize, labelformat=simple, figurename=Fig., labelsep=space]{caption}
\usepackage[flushleft]{threeparttable}
\usepackage{graphicx}
\usepackage[backend=biber, style=apa]{biblatex}
\DeclareLanguageMapping{english}{english-apa}
\usepackage{url}
\usepackage{hyperref}
\usepackage[toc,page]{appendix}
\usepackage{longtable}
\usepackage{chngcntr}
\usepackage{wrapfig}
\usepackage{float}
\usepackage[font=footnotesize]{subcaption}
\usepackage{tabto}

\newcommand{\nocontentsline}[3]{}
\newcommand{\tocless}[2]{\bgroup\let\addcontentsline=\nocontentsline#1{#2}\egroup}

\bibliography{mcf_bib.bib}
\IfFileExists{upquote.sty}{\usepackage{upquote}}{}
\begin{document}

\pagenumbering{Roman}

%\includepdf[pagecommand={\begin{tikzpicture}[remember picture, overlay]\node at (current page.center) {overlayed text};\end{tikzpicture}}]{Deckblatt_Master17.pdf}

\title {Heterogeneous Effects of Unconventional Monetary Policy on Loan Demand and Supply. Insights from the Bank Lending Survey}
\author {Martin Guth\textsuperscript{1}*}
\date {}
\maketitle 

\section*{Abstract}
This paper analyzes the bank lending channel and the heterogeneous effects on the euro area, providing evidence that the channel is indeed working. The analysis of the transmission mechanism is based on structural impulse responses to an unconventional monetary policy shock on bank loans. The Bank Lending Survey (BLS) is exploited in order to get insights on developments of loan demand and supply. The contribution of this paper is to use country-specific data to analyze the consequences of unconventional monetary policy, instead of taking an aggregate stance by using euro area data. This approach provides a deeper understanding of the bank lending channel and its effects. That is, an expansionary monetary policy shock leads to an increase in loan demand, supply and output growth. A small north-south-disparity between the countries can be observed.
\\
\\
\textit{Keywords:} Monetary Policy, ECB, BLS, Loan Demand, Loan Supply
\\
\textit{JEL classifications:} C4, E0, E4, E5
\\\\\\\\\\\\\\\\\\\\\\\\
{\footnotesize \textsuperscript{1}Oesterreichische Nationalbank (OeNB), Supervision Policy, Regulation and Strategy Division, Otto-Wagner-Platz 3, 1090 Vienna, Austria (martin.guth@oenb.at)\par}
{\footnotesize \noindent *Opinions expressed by the author do not necessarily reflect the official viewpoint of the Oesterreichische Nationalbank or of the Eurosystem.\par}

\newpage

\clearpage 
\pagenumbering{arabic}
\newpage

\section{Introduction}
\setcounter{page}{1}

The last years have been turbulent times for the European Central Bank (ECB) and the other European national banks. Starting with aggravated financial conditions due to the emerging US sub-prime crisis in August 2007, which was followed by the collapse of Lehman Brothers on the 15th September 2008 and the global financial crisis that unfolded after it. 

The negative impact of the crisis on the banking system culminated in extraordinary measures banks have taken in the upcoming weeks and months after the collapse. Loan policies have been revamped and loan conditions have been tightened to stabilize balance sheets. 

The first shocks to the financial system have been followed by increasingly nervous markets anticipating a Greek sovereign default in early 2010, with a risk of contagion to other European countries. As the sovereign debt crisis hit Italy and Spain in mid 2011, the euro area banking system faced serious problems regarding tumbling balance sheets and a dried-up interbank market \parencite{philippine:2013}. The close link between financial and non-financial markets led to real effects on the European economy. GDP growth in the euro area slowed down by 3.8\%\footnote{Average calculated for the sample countries in this paper. Values for the GDP growth taken from \textcite{oecd:2015}.} during the first years of the crisis compared to the growth level of 2007. This comes at no surprise, as lending from banks accounts for up to 85\% of the sources of external finance to corporations and households. Hence, the tightened loan supply had a negative second round effect on output growth \parencite{ecb:2009}.

When transmission channels of conventional monetary instruments are impaired by market stress, the use of unconventional policies can help open up new channels to stimulate bank lending and thus output \parencite{albertazzi:2016}. In early 2008, the first enhanced credit support policies have been issued, which were designed to provide more liquidity to banks in an easier fashion. The policies included fixed-rate full allotment, giving banks unlimited access to central bank liquidity, extended maturities on liquidity provisions and the extension of eligible collaterals. Additionally, several long-term refinancing operations (LTROs) had been released with a maturity date ranging between 6 to 48 months. Those helped banks to stabilize their balance sheets in the short-term and strengthen their liquidity position in the long-term. As bank liquidity is not the only concern of the ECB, an asset purchase program has been launched to further boost economic activity and drive up inflation via increased capital flows at the zero lower bound \parencite{trichet:2009}.

Even though the ECB deployed many non-standard measures to counter the negative effects on the financial system, this paper focuses on policies which had a direct effect on banks and the provision of loans. It is thus consistent with the goals of the ECB “[...] to maintain the availability of credit for households and companies at accessible rates” \parencite[p. 12]{trichet:2009}. In times where financial frictions are increased, i.e. the financial crisis from 2008 onward, the availability of loans is decreased due to more stringent terms and conditions. The scientific literature has already shown that in such times, (unconventional) monetary policy has a higher impact on GDP through the credit channel, as excess liquidity helps banks to soften their lending conditions \parencite{bernanke:1995, kashyap:2000}. 

Concluding the outline, the research questions can be formulated as follows: \textit{How effective was the ECB with their liquidity providing policies? Was the bank-lending channel operational during the crisis? If yes, what are the different heterogeneous country-level effects of these euro-wide policies on loan demand and supply?}

In order to identify the credit channel, the paper makes use of the Bank Lending Survey (BLS) from the ECB to get an insight into the demand and supply of loans from and to enterprises. In contrast to the majority of scientific literature on this topic, which investigates the aggregate effects, this paper investigates the country-level impulse response functions in order to get a deeper understanding of the heterogeneous effects of a unified monetary policy shock. In order to do so, a vector autoregressive (VAR) model is deployed which will be estimated by Bayesian methods.

The findings of this paper show a clear indication of country-specific effects from an unconventional monetary policy shock. Following such a shock, the loan supply condition loosens up for around 3\% for several months. However, Germany, Greece and Italy show an apprehensive behavior regarding their lending conditions as the responses are delayed for 5-6 months. A similar delay can be observed for five out of seven countries for the development of loan demand, for which the median peaks at 2\% to 3\%. In situations where interest levels are low, investments indeed react slowly and this could be correlated to the delayed credit demand responses \parencite{praet:2017}. The effects on GDP growth reveal that the broad credit channel is indeed working, even though the transmission channels are impaired. These findings show a clearer north-south-disparity as Austria, Belgium, Germany and Italy show significant positive responses and Greece, Portugal and Spain show small or no effects.

In order to test the robustness of the findings, a wide variety of checks will be deployed. First, the lag length will be extended from two to three and four lags. Second, the Cholesky ordering will be changed in a way that the two loan variables will be ordered below the shock variable. This would indicate that loan demand and supply would react within the same month as the monetary policy shock is deployed, and that the ECB does not base their policy decisions on the latest Bank Lending Survey. Third, the shock variable will be changed to the spread between the main refinancing operations (MRO) and Euro OverNight Index Average (EONIA) rate as proposed by \textcite{albertazzi:2016} and \textcite{lenza:2010}. The authors show that the spread is correlated with the excess liquidity of the ECB and thus should be even more useful in capturing credit enhancing policies. Overall, the results of those tests confirm the findings of the base model.

%Sign restrictions here

The remainder of the paper is organized as follows: The next section provides a short review of the literature. Section~\ref{sec:data} introduces the data, some descriptive analysis and the methodology. Section~\ref{sec:results} presents the results. Section~\ref{sec:con} summarizes and concludes the paper.

\section{Literature Review}

This section presents a short overview of the three main topics discussed in this paper. We start with an explanation of the basic banking channels at work between the monetary policy of the ECB and the banks. Subsequently, a brief outline of the ECB's monetary policy course from 2008 onward is given. Finally, other papers analyzing the effects of monetary policy on the development of loans, especially with the usage of the BLS, are presented.

%BANK LENDING CHANELL
The broad credit channel, as hypothesized first by \textcite{bernanke:1988} and later on formulated in detail by \textcite{bernanke:1995} and \textcite{kashyap:2000}, arises due to frictions within the financial system. More precisely, it refers to the role of financial institutions to transmit monetary policy to the economy via loans and other financial products. Such frictions could be asymmetric information between lenders and borrowers or a severe disturbance in the market. Both scenarios would lead monetary financial institutions to act less risky and retain their liquidity instead of investing in other risky products or handing out loans to non-financial corporations and households. This leads to a situation where banks and the rest of the real economy is short on liquidity and thus investment can be hampered. However, the credit channel is not a transmission channel on its own, but a amplification of the effects of monetary policy. In theory, central banks indirectly effect the external finance premium of banks which is defined by the difference in costs of the generation of external and internal funds. This effect can be split up in two sub-channels: the balance sheet channel and the narrow credit channel.

The balance sheet channel stems from the fact that a strong financial position, in the form of liquid assets or valuable illiquid assets like real estates, can help to reduce the conflicts between lender and borrower. Hence, the balance sheet of a corporation and its fluctuations influence the external financial premium. A negative monetary policy shock influences the balance sheet via increased interest expenses on debts and declining asset prices. 

The more interesting channel for this paper is the narrow credit channel. This channels represents the more direct effects of the broad credit channel, such that monetary policy influences the supply of loans via in- or decreased liquidity supply. In a rather straightforward mechanism, banks can overcome the frictions of a dysfunctional market by lending money from central banks. As \textcite{allen:2004} show, financial institutions are important for the provision of loans in the euro area. Thus a reduction of possible credit supply by financial institutions incurs costs on the borrowers as they have to search for new sources of liquidity.

More recently, \textcite{nicolo:2010} proposed a third credit channel in the form of increased risk taking by banks due to the low interest rate environment. Their observations are based on the fact that banks reallocate their balance sheets from safe to risky assets to gain higher yields. Even though this channel has no direct effect on bank lending, the risk appetite of financial intermediaries has a positive impact on the overall credit supply \parencite{naceur:2017}.

%The authors conclude that such an impact varries across countries and banking systems as higher capitalized banks will take up more risk. 
%The channel splits further into three sub-channels in which increased risk taking is precipitated via substition of safe low yield assets to more risikier high yield assets, taking up more long-term assets from saving institutions and procyclical leverage ratios, as banks try to hold their leverage ratio constant by changing the composition of assets during boom and bust phases.

A credit enhancing monetary policy helps banks to acquire money in the market more easily. With more liquidity flowing in and a strong balance sheet, banks can act more venturesome and soften their loan terms and conditions. Through this mechanism, real economic activity and thus inflation can be boosted.  

%MONETARY POLICY COURSE
Now one could ask the question if the ECB followed the theory and deployed such enhanced credit support policies. Looking at the course of the European central bank's monetary policy actions, we can see that indeed the measures have been liquidity providing and thus credit enhancing. This fact comes at no surprise as the ECB always stated that the provision of credits to households and corporations is one of the main priorities \parencite{trichet:2009}. The implemented policies are all well-tailored for the bank-centered financial system of Europe \parencite{philippine:2013}. Comparing the actions taken by the ECB, the FED, the Bank of England and the Bank of Japan after the eruption of the crisis, \textcite{fawley:2013} conclude that the European monetary policies can be considered "pure" (p. 55) quantitative easing as they target central bank reserves and allow for a wide range of collaterals.  

One of the first policies seen after the bankruptcy of the Lehman Brothers was the change from variable-rate tenders with a bidding process to fixed-rate tenders and full-allotment (FRFA) for all future lending operations.\footnote{A detailed summary of the actions taken by the ECB can be seen in \textcite{fawley:2013}, \textcite{fratzscher:2014} and \textcite{philippine:2013}. This section presents the most prominent points out of those papers to give a summarizing overview.} Which means that banks could now lend as much as they want from the ECB at the main refinancing operations rate. At the same time, the ECB started to widen the range of possible collaterals for the lending activities of the banks. Thereby banks could refinance more illiquid assets in their balance sheets. This combination helped banks to access short-term liquidity to relief the stress of their balance sheets. On the side of conventional monetary policy, the main interest rate was cut by 50 basis points from 4.25\% to 3.75\% in October 2008 and was since then slowly lowered to the zero lower bound, which was hit on the 16th of March, 2016.

Now that short-term liquidity was secured, the Governing Council of the ECB turned towards the medium- and long-term lending. The conventional longer-term refinancing operations (LTROs) have a maturity of 2 weeks to 3 months. Early in 2008, the maturity has been increased to 6 months, with several operations being carried out between 2008 and 2011. From 2009 onward till 2011, four 12-month refinancing operations have been carried out. With the sovereign debt crisis hitting the financial system in the fall of 2011, the ECB pushed the maturity for the upcoming LTROs to 36 months.\footnote{Also known as very longer-term refinancing operations (VLTROs)} Those two 3-year operations allotted around €1 trillion in total into the system. In 2014 and 2016, targeted longer-term refinancing operations (TLTROs), with a maturity of 48 months, had been issued. The amount banks could borrow were linked to the amount of lending to non-financial corporations and households each bank had in its books. The second round of the TLTROs had a new implementation of linked interest rates to the lending pattern of banks. The more loans a bank would issue for investments or consumer credit, the more attractive the interest rate for the lending operation would become. With this new setup, the ECB tried to direct the flow of funds directly to the real economy, as banks could have the incentive to take the excess liquidity and invest into high-risk products boosting their revenue and not the economy. All those extensions did not only keep money market rates low, but also strengthen the position of the ECB as intermediary between the banks and the money market, which clearly helped the financial institutions to strengthen their overall condition.

As already mentioned, this paper focuses on the credit enhancing policies. Hence, programs such as the currency swap agreements, the Outright Monetary Transactions (OMT) or the Securities Markets Programme (SMP), which where set in place to relief tensions in certain markets that impaired the transmission mechanism of the monetary policy actions, or the covered bond purchase programme (CBPP), aimed at boosting economic activity and inflation, are not covered.

%BLS USAGE
After setting up the theoretical background between the monetary policy transmission channels and the actual monetary policies, the question arises how to identify the channels and the actions taken. In order to properly identify the narrow credit channel, a database is needed that provides information on the supply and demand of bank loans. Before 2003 this gap could not be filled in a satisfying way, but since then, the ECB conducted the Bank Lending Survey with detailed questions for banks in the euro area and thereby asses the lending situation. The second problem arises from the question how to quantify the credit enhancing policies. The scientific community came to the conclusion that the Euro OverNight Index Average (EONIA) rate would be a credible tool for assessing the monetary policies of the ECB.\footnote{A more detailed explanation of the BLS and the EONIA rate can be found in section~\ref{sec:var}.}

Looking at related literature, \textcite{ciccarelli:2013} investigate the heterogeneous effects of a monetary policy shock on countries which are under sovereign stress and other eurozone countries.\footnote{The group of countries under stress is composed of Greece, Italy, Ireland, Portugal and Spain. The other eurozone countries are Austria, Belgium, Finland, France, Germany, Luxembourg and the Netherlands} For the identification process, they rely on the answers from the BLS and the EONIA rate as shock parameter. Their results point out that the liquidity provision of the ECB has helped the distressed countries to loosen their loan standards, thereby increasing the amount of loans supplied by banks and hence increase the GDP growth significantly. 
The paper by \textcite{couaillier:2015} focuses on the effects of conventional and unconventional monetary policy on the terms and conditions of loan contracts. He also relies on the Bank Lending Survey and a bayesian approach to overcome the short sample of the survey. The findings show that a increase in a monetary policy rate, here the deposit facility rate, has an negative impact on the terms and conditions and thus on the credit supply.
\textcite{darracq:2015} use the BLS in a new way to identify the impact of the 3-year LTROs. The authors conclude with their VAR analysis that the longer-term refinancing operation had an positive impact in the short to medium term on GPD, inflation and loan supply to enterprises. Additionally, the LTROs helped banks to lower their leverage and to increase their liquidity and capital buffers.
\textcite{ozsahin:2016} focuses his research on the characteristics and condition of banks and how those influence the transmission of monetary policy on output. The identification of the credit variables is also based on the BLS and identical to \textcite{ciccarelli:2015}. The author argues that stricter bank regulation, in terms of bank capitalization and stable funding base, has a positive effect on the transmission of monetary policy through the bank lending channel.
The last paper to mention is from \textcite{boeckx:2017}. The authors also use the BLS to identify demand and supply of loans, but do not model a pure monetary policy shock and instead construct a balance sheet shock. For the shock only the balance sheet exposure to the banking system is used, as this relates to volume of credit to the non-financial corporations and households. Thereby capturing all actions taken by the ECB such as the longer-term refinancing operations and the covered bond purchase program. \textcite{boeckx:2017} conclude that an expanding balance sheet has a positive effect on GDP growth, prices and lending.

\section{Data and Methodology}
\label{sec:data}

The research focuses on the heterogeneous effects of an unconventional monetary policy shock on loan demand and supply. In order to gain useful insights from the Bayesian VAR model, one needs to carefully identify the shock of interest, which isolates the liquidity providing measures of the ECB into one variable. The possible candidates and alternatives for the robustness checks are described below. The sample consists out for seven countries, namely Austria, Belgium, Germany, Greece, Italy, Portugal and Spain. The limitation to those countries comes from the Bank Lending Survey, as countries can report their surveys in different indices. Those indices, however, are not comparable to each other. The usage of the net percentage index thus yields the biggest country sample. Furthermore, these countries report data within the same time span.

\subsection{Variable choice}
\label{sec:var}

The variables can be grouped into three thematic blocks, which will be discussed thoroughly below. The monetary policy block is used to control for the shock, but also to approximate central bank behavior. The Bank Lending Survey block consits of the survey indicators regarding the development of loans in the euro area. The third block captures the generel macroeconomic setting. The data spans from January, 2008 to March, 2017. In order to increase the number of observations the model uses monthly data points. This yielded the problem that GDP growth, loan demand and loan supply had to be interpolated from a quarterly to a monthly time frame. The interpolation itself is described in the respective sections below. \newline

\begin{figure}[h!]
\centering
 \resizebox{0.8\linewidth}{0.7\linewidth}{
\begin{knitrout}
\definecolor{shadecolor}{rgb}{0.969, 0.969, 0.969}\color{fgcolor}
\includegraphics[width=\maxwidth]{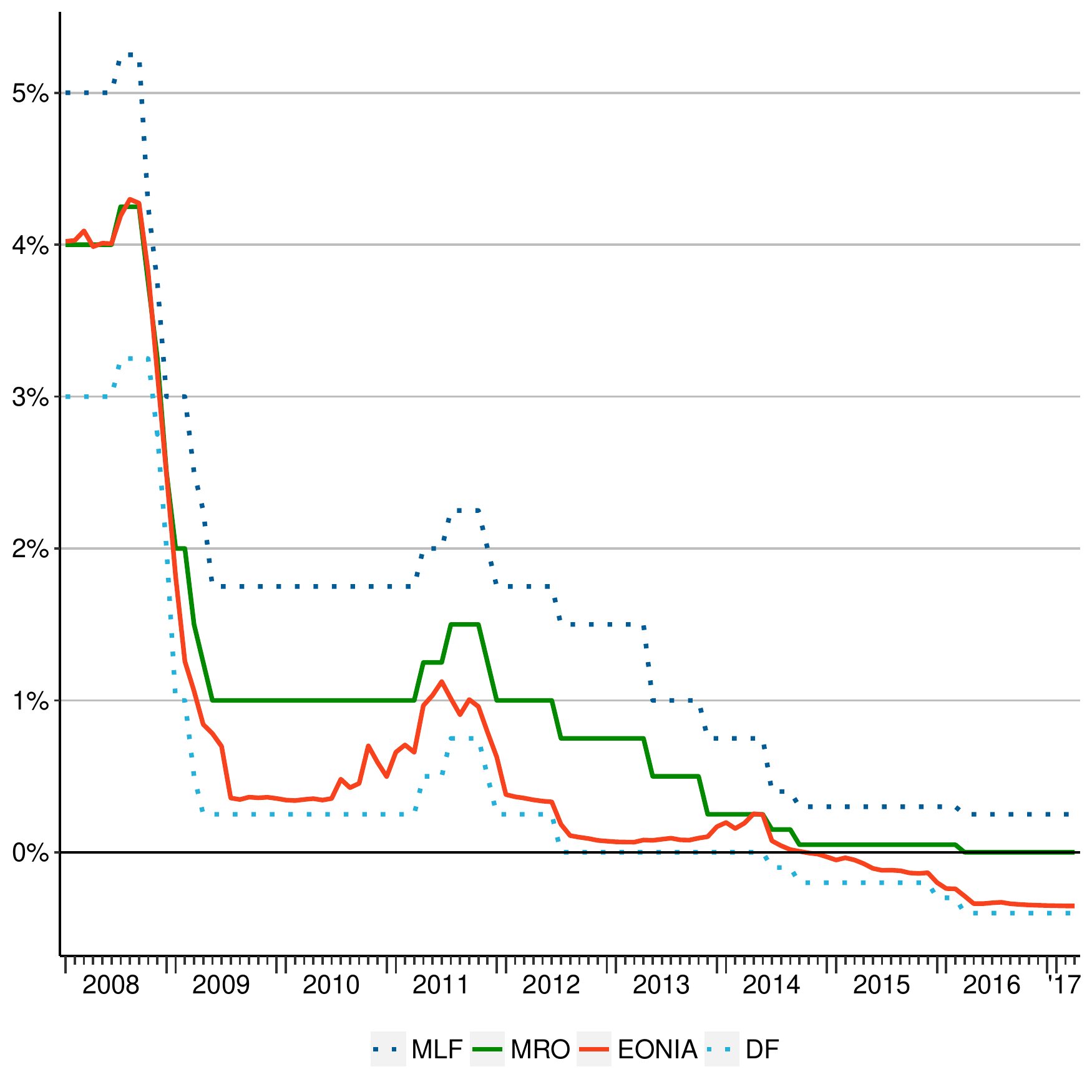} 

\end{knitrout}
}
\caption{Monetary Policy Rates of the ECB. \textit{Note}: The \textit{dark blue dotted line} represent the marginal lending facility (MLF) rate at which banks can obtain overnight liquidity and the \textit{light blue dotted line} represents the deposit facility (DF) rate at which counterparties can deposit money at the central bank. Those two lines make up a corridor for the main policy rate of the ECB, the main refinancing operations (MRO) rate, here drawn as the \textit{solid green line} and for the Euro OverNight Index Average (EONIA) rate which is the weighted average over all overnight interbank lending operations, here drawn as the \textit{solid red line}.}
\label{fig:mprates}
\end{figure}

\noindent \textbf{Monetary policy block} \newline
The ECB makes use of three different interest rates to display its stance on monetary policy. The marginal lending facility (MLF) rate and the deposit facility (DF) rate are important for the access to overnight liquidity for banks. Those two indicators represent the ceiling and the floor for the most prominent monetary policy indicator, the main refinancing operations rate, at which banks can lend money from the ECB for one week. In figure~\ref{fig:mprates}, the MLF-DF corridor is plotted between 2008 and 2017. A major downward shift can be seen between the end of 2008 and the middle of 2009, pushing the MRO rate to 1\%. The (near) zero lower bound was targeted from 2014 onwards. All these decisions are based on the provision of easy liquidity to monetary financial institutions and at the same time prevent them from gaining interest income via deposits at the ECB.

The most common interest rate to control for unconventional monetary policy is the Euro OverNight Index Average (EONIA) rate \parencite{ciccarelli:2013}. It is the weighted average over all overnight interbank lending operations between 28 panel banks \parencite{ecb:2018}. Hence, by definition the EONIA rate also moves in the corridor between the MLF and DF rate. Until the start of the crisis, the overnight index rate was seemingly identical to the MRO rate. This fact changed with the start of the liquidity providing operations of the ECB and thus the EONIA rate dropped below the MRO rate. ECB's Governing Council used the deposits facility rate to drag the EONIA more to the floor, such that the monetary policy decisions could be properly transmitted to the financial system \parencite{trichet:2009}.

However, there are other authors which argue that the spread between the MRO and the EONIA rate would be the better way to capture the liquidity providing measures.\footnote{See \textcite{albertazzi:2016}, \textcite{boeckx:2017} and \textcite{lenza:2010}} They show that unconventional monetary policy can trigger a liquidity effect which drives a wedge between the MRO and EONIA rate. Further, this indicator is correlated with excess liquidity of the central banks in the euro area during their longer-term refinancing operations. However, the spread will not be included in the base model, but it will be used as shock variable in one of the robustness checks.\\

\noindent \newline \textbf{Bank Lending Survey block} \newline
The Bank Lending Survey is carried out quarterly by the ECB and provides information on the lending situation of banks in the euro area. The senior loan officers of 140 sample banks from 19 countries are asked about their opinion on questions related to the terms and conditions of their loan contracts, developments of loan demand or the change of lending standards \parencite{koehler:2016}. All questions are asked backward looking for the last quarter and forward looking for the upcoming quarter. This paper uses the backward looking questions as they resemble the actual lending situation and not a prediction of it.

The possible answers to the survey range from "eased considerably" to "tightened considerably" or "increased considerably" to "decreased considerably" and are thus qualitative answers. The BLS database provides multiple transformations, however, as many other scientific papers, this paper too follows \textcite{lown:2006} by using the BLS variables as net percentages. These values are calculated from the percentage difference of all banks per country reporting a positive and vice versa a negative development.

The number of chosen countries depends on the availability of data regarding the BLS. Even though 19 countries take part in the survey, not all report their results with the same definitions or in the same time span. Thus the resulting list of countries shrinks to seven in total, namely Austria, Belgium, Germany, Greece, Italy, Portugal and Spain.

For the analysis, the paper makes use of two main questions from the survey, i.e. "Over the past three months, how have your bank’s credit standards as applied to the approval of loans or credit lines to enterprises changed?" which represents the development of loan supply and "Over the past three months […] how has the demand for loans or credit lines to enterprises changed at your bank?" which represents the development of loan demand. As the time frame is limited, the variables are only for the loan situation against non-financial corporations.\footnote{The BLS also includes information on households and mortgages.} 

\begin{figure}[h!]
\centering
 \resizebox{1\linewidth}{!}{
\begin{knitrout}
\definecolor{shadecolor}{rgb}{0.969, 0.969, 0.969}\color{fgcolor}
\includegraphics[width=\maxwidth]{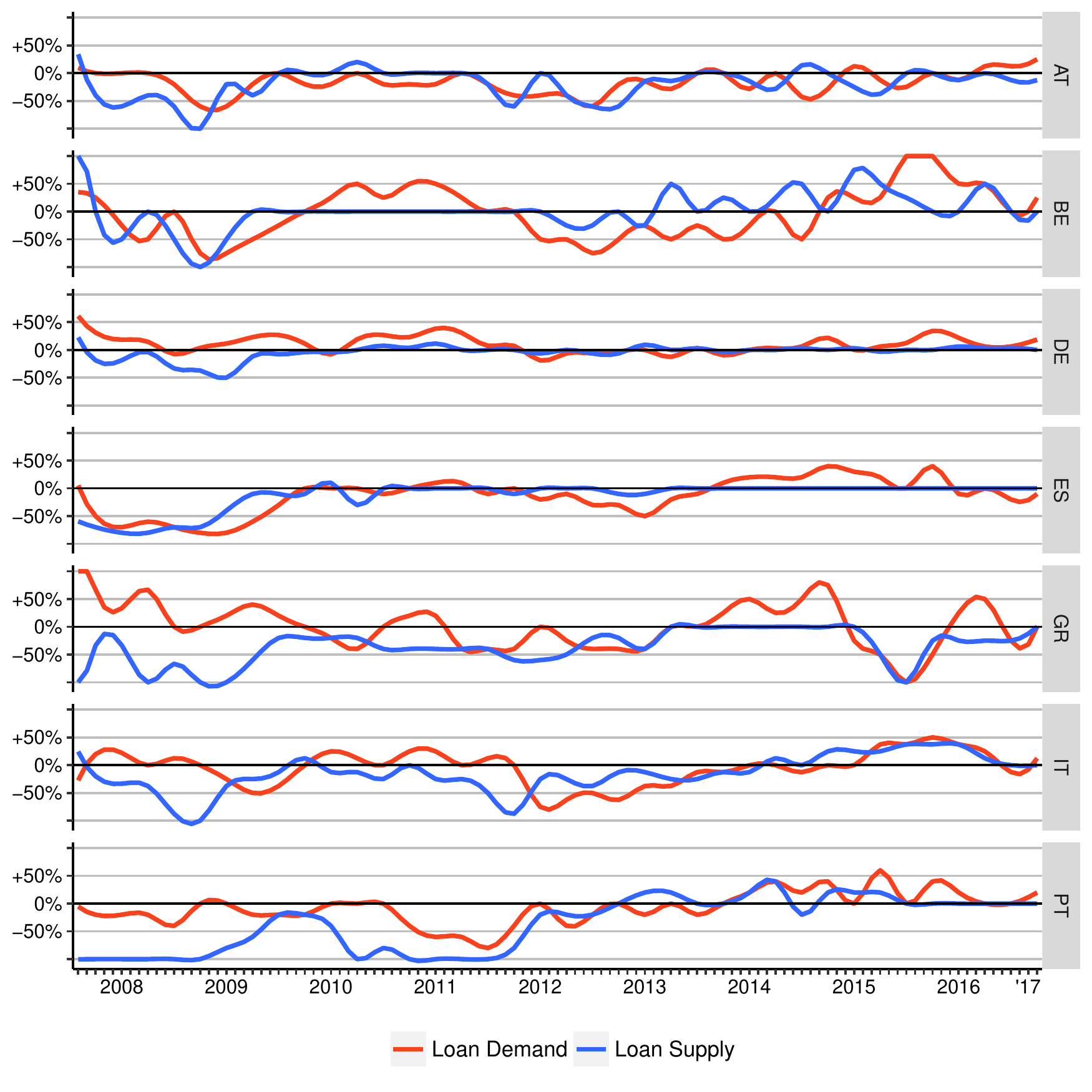} 

\end{knitrout}
}
\caption{Time series of enterprise credit variables for all available countries. \textit{Note}: The \textit{solid red line} represents the loan demand taken from the Bank Lending Survey (BLS) as reported by the local loan officers for the last three months. The \textit{solid blue line} is the loan supply also taken from the BLS for the last three months. Normally, the supply is reported as the tightening of credit standards. In order to give a more coherent picture, the loan supply is inverted, such that an a positive value on the chart represents an increase in credit supply. Both values are reported in net percentages, which is the percentage difference between banks reporting in overall increase to those banks reporting an overall decrease in loan demand (supply).}
\label{fig:bls}
\end{figure}

Figure~\ref{fig:bls} depicts the developments of loan demand and supply on country level. One can see that the changes in the variables are highly heterogeneous across the countries. There are time periods where the indicators yield a zero percentage change. This could be due to the fact that the aggregation method used by the ECB could have sterilizing effects on the developments. As the BLS does not report the bank-specific answers to the survey, a country could be split fifty/fifty between loosening (increasing) and tightening (decreasing) loan developments. Normally, the BLS gathers information on the loan supply condition in such a way that an increase in the data would mean harsher loan condition and less supply. In order to give a more comprehensible picture, the data is inverted. As \textcite{bondt:2010} or \textcite{giovane:2011} have shown, the usage of questions concerning the credit standards is a legitimate instrument to capture the credit availability in the euro area. A contractive shift, in both the supply and demand curve of bank lending, will lead to a reduction of total credit volume. The strength of each shift decides the resulting loan interest rate \parencite{ecb:2009}. Therefore, this paper uses both loan demand and supply in the model as both variables are important in explaining the lending situation in the euro area.

Both BLS variables are interpolated to get from a quarterly to a monthly time frame. The cubic spline method is used as there is no clear indicator to proxy the development of credit demand or supply.\footnote{A Chow-Lin based interpolation as in \textcite{peersman:2011} cannot be conducted as the development of the index on the adjusted loans to euro area non-financial corporations does not correlate with the BLS data on loan supply.}\\

\noindent \textbf{Macroeconomic block} \newline
This block includes macroeconomic control variables of different forms. Nearly every variable in this block is available on the country level, which means that they capture different heterogeneous effects. Output growth and prices control for the general macroeconomic situation in each country. \textcite{jimenez:2012} show that "[...] under tighter macro conditions [..] a capital crunch begets a credit crunch." (p. 2303). The period of the financial crisis can clearly be seen as a tighter macro condition, which in turn is captured by a dropping GDP growth. It is taken from Eurostat at market prices, seasonally adjusted and included in terms of the quarter-on-quarter growth rate. Since GDP is not available on a monthly basis, a Chow-Lin interpolation is applied with industrial production serving as proxy. 

The ECB uses the Harmonised Index of Consumer Prices (HICP) for their inflation goals. However, the consumer price index including energy prices is driven by rising (sinking) commodity prices, i.e. oil prices. Especially the consumer prices for liquid fuels are directly linked to oil and make up nearly the half of the energy component of the HICP \parencite{ecbhicp:2016}. Thus, this paper follows \textcite{couaillier:2015} and uses the core HICP excluding energy and unprocessed foods as monthly rate of change. Thereby avoiding an upwards (downward) bias and getting closer to the more broader inflation rate of the GDP deflator \parencite{alcidi:2016}.

The Composite Indicator of Systemic Stress (CISS) reveals turmoil in the financial markets and systemic risk. It is taken from the ECB's statistical data warehouse and developed by \textcite{hollo:2012}. It is composed of money market, bond market, equity market, financial intermediaries and foreign exchange market indicators. As \textcite{boeckx:2017} points out, the indicator is helpful in explaining monetary policy decision based on financial stress and uncertainty.

After the macroeconomic and financial control variables, the heterogeneous banking system also needs control variables. For this approach, the paper follows \textcite{ciccarelli:2013} and includes two control variables.\footnote{\textcite{ciccarelli:2013} actually uses three variables to capture the heterogeneous effects. However, long-term liquidity provided by the Eurosystem to the banking sector on a country-level could not be obtained from a public database.} First, the long-term government bond yields which proxy for the respective country risk. Second, the interbank liquidity to account for the stability of bank funding which does not rely on ECB money. However, the transaction volumes are not publicly available and are thus proxied by the deposit liabilities vis-a-vis euro area monetary financial institutions (MFI). A bank's deposit from other banks must come from the interbank market and can does be seen as interbank transaction volumes.

The last variable in this block is the Euro Stoxx 50 index, which covers the development of the 50 biggest publicly traded companies in the Eurozone.

\subsection{Econometric Model}
\label{sec:econ}

The econometric model adopted is a vector autoregressive (VAR) model and follows the specification in \textcite{huber:2017}. In regression form, the VAR model is given by,
\begin{equation}
Y_t = AX_t + \epsilon_t, \text{ for } t=1,\dots,T,
\end{equation}
where $Y_t$ is a $m$-dimensional vector of endogenous variables, $A = (A_1,\dots,A_p)$ is a $m \times mp$-dimensional coefficient matrix, where $p$ denotes the lags,\footnote{For the base model two lags as proposed in \textcite{ciccarelli:2015} are used.} $X_t = (Y_{t-1},\dots,Y_{t-p})$ is a $mp$-dimensional vector of lagged endogenous variables and $\epsilon_t$ is a normally distributed error term with zero mean and a variance-covariance matrix $\Sigma_t$, such that,
\begin{equation}
\Sigma_t = H^{-1}S(H^{-1})'
\end{equation}
$H^{-1}$ denotes a lower triangular matrix with unit diagonal and $S =\\diag(s_1,...,s_m)$ is a diagonal matrix of variances.

The used time frame starts with the first month of 2008 and spans till the third month of 2017 resulting in $t = 111$ time points. As described in section~\ref{sec:var}, the $Y_t$ vector is composed of the MRO rate, EONIA rate, loan demand, loan supply, GDP growth, HICP index, CISS indicator,  long-term government bond yields, MFI deposits and Euro Stoxx 50 index. This yields 52 euro area and country-specific variables. Thus the total number of autoregressive parameter $k = m(mp)$ exceeds the number of time points $t$, which calls for a Bayesian estimation approach.

The hierarchical prior specification follows \textcite{huber:2017}, which relies on a normal-gamma shrinkage prior. In order to handle the large number of parameters, \textcite{huber:2017} impose a Gaussian prior on each coefficient $\alpha = vec(A)$, such that,
\begin{equation}
\alpha_i|\psi_i \sim \mathcal{N}(0,2/\lambda_{\psi}^2 \psi_i), \psi_i \sim \mathcal{G}(\vartheta_\psi,\vartheta_\psi), i = 1,\dots,k.
\end{equation}
This prior setup has the handy feature of using a lag-specific global shrinkage parameter $\lambda_{\psi j}^2$ for $j = 1,\dots,p$, such that,
\begin{equation}
\lambda^2_{\psi i} = \prod_{j=1}^p \zeta_j,~\zeta_j \sim \mathcal{G}(d_j,l_j),
\end{equation}
which pushes all elements in the coefficient matrix, and especially the coefficients with higher lags, strongly towards zero. Additionally, the prior includes a local shrinkage parameter $\psi_i$ that allows for non-zero coefficients even with a strong global shrinkage parameter. The mean and the variance of $\psi_i$'s gamma distribution $\vartheta_\psi$ is a hyperparameter chosen by the researcher. As the authors describe in their paper, a small $\vartheta_\psi$ places more mass on the zero mean of $\alpha_i|\psi_i$ while the tails become heavier. Thus, such a hyperparameter setup would still provide enough flexibility for the model to make use of the data characteristics.

The actual estimation of the econometric model is achieved by a Markov Chain Monte Carlo (MCMC) algorithm described in \textcite{huber:2017}.

\subsection{Identification}
\label{sec:ident}

As already mentioned, the aim of this paper is to evaluate the impact of an unconventional monetary policy shock on loan demand and supply and, as a second round effect, on output and inflation. In order to identify the shock in the VAR model a Cholesky decomposition with a specific variable ordering is employed. As the ordering of the variables is crucial for structural inference, this paper draws from the ECB's two-pillar approach on monetary policy strategy and therein especially the second pillar. With the maintenance of price stability as primary objective, this pillar evoles around an assesment of the economic and finacial conditions and their effect on prices. The analysis of those indicators serves as guidance for upcoming monetary policy decisions \parencite{ecb:2000}. For the Cholesky ordering, this implies that prices, output growth, the financial stress indicator and both, loan demand and supply, are ordered before the shock variable.\footnote{Placing loan supply before demand has virtually no effect on the results.} Every quantity above the policy variable does not change contemporaneously with the shock. This paper also includes the main refinancing operations rate before the shock in order to capture the zero lower bound. After the EONIA rate, the fast moving variables are placed in the following order: long-term government bond yields, the MFI deposits and the Euro Stoxx 50 index. Especially the MFI deposits, which capture the heterogeneous interbank funding situation, are expected to influence the demand for ECB liquidity and thus the main policy rate.

Hence, the 52 variables in $Y_{t}$ are in the following order: HICP index, GDP growth, CISS indicator, loan demand, loan supply, MRO rate, EONIA rate, long-term government bond yields, MFI deposits and Euro Stoxx 50 index.\footnote{All variables are for the full country sample except the MRO rate, EONIA rate and the Euro Stoxx 50 indicator.}

\section{Impact of Credit Support Policies}
\label{sec:results}

All impacts reported in this section are normalized by the size of the shock and scaled to 25 basis points. The normalization alters the y-axis to be in percentage points and makes the plots comparable. The shock is inverted to resemble a expansionary monetary policy decision, which means that the EONIA rate decreases. Each plot displays the median response along the grey shaded area which represents the 16th and 84th Bayesian credible interval.

\begin{figure}[h!]
  \begin{subfigure}{.12\textwidth}

  \caption*{}
  \label{fig:dummy3}
  \end{subfigure}
  \begin{subfigure}{.24\textwidth}
  \caption{Austria}
  \label{fig:AT_Supply}

\includegraphics[width=\maxwidth]{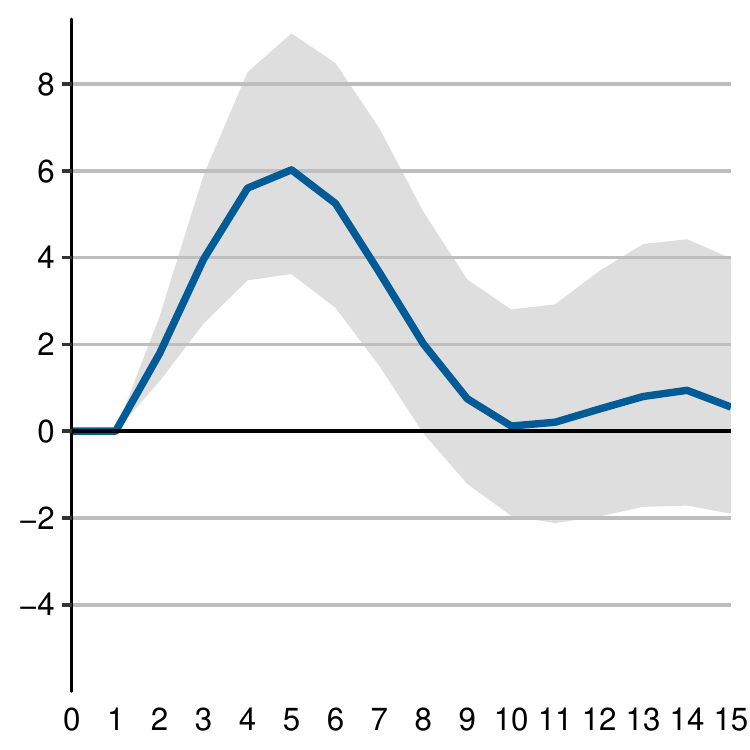} 

  \end{subfigure}
  \begin{subfigure}{.24\textwidth}
  \caption{Belgium}
  \label{fig:BE_Supply}

\includegraphics[width=\maxwidth]{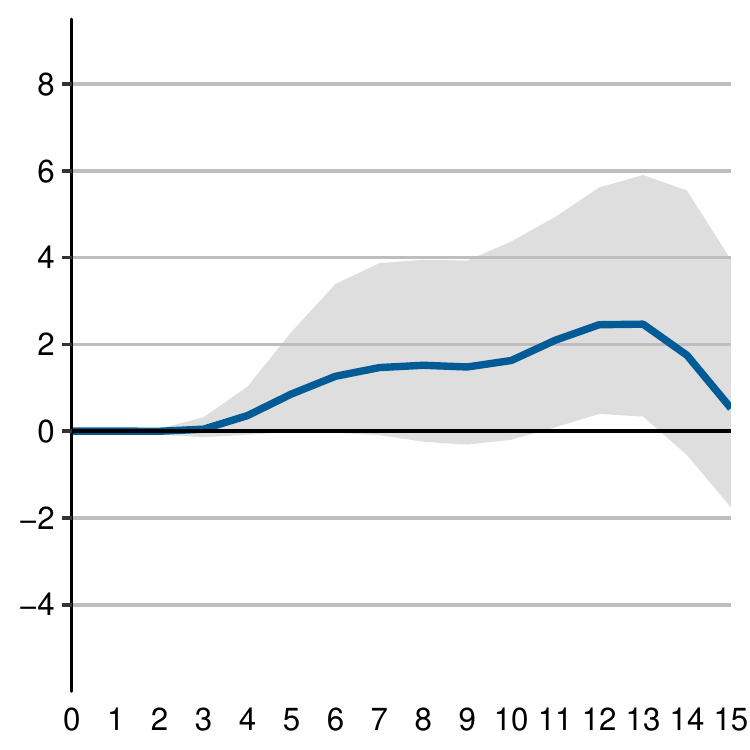} 

  \end{subfigure}
  \begin{subfigure}{.24\textwidth}
  \caption{Germany}
  \label{fig:DE_Supply}

\includegraphics[width=\maxwidth]{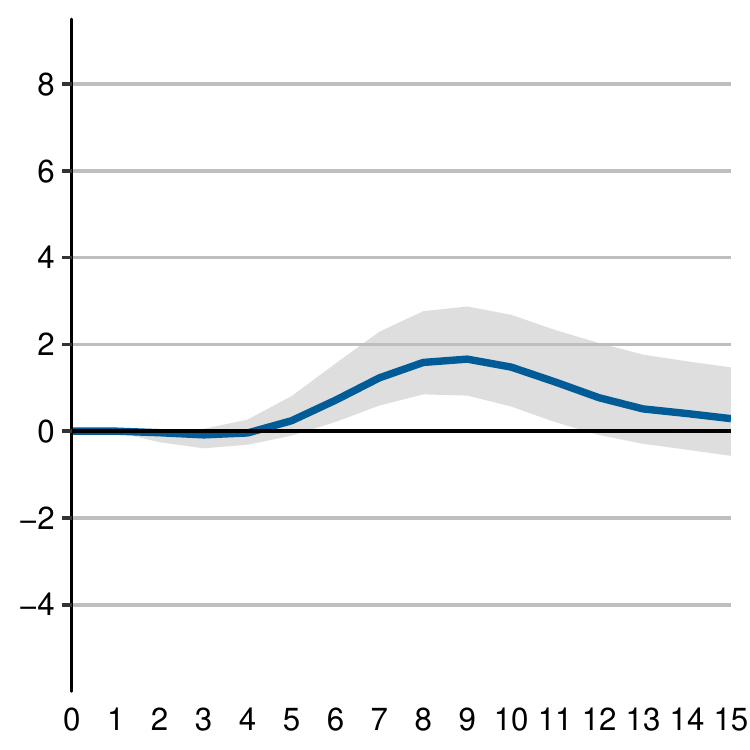} 

  \end{subfigure}
  \begin{subfigure}{.12\textwidth}

  \caption*{}
  \label{fig:dummy4}
  \end{subfigure}
  
  \vspace{0.3cm}
  
  \begin{subfigure}{.24\textwidth}
  \caption{Greece}
  \label{fig:GR_Supply}

\includegraphics[width=\maxwidth]{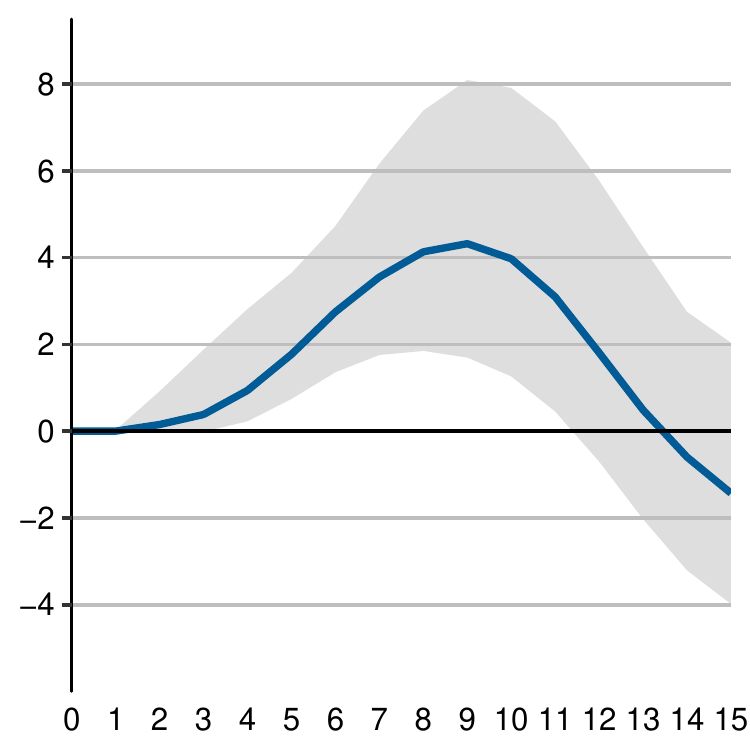} 

  \end{subfigure}
    \begin{subfigure}{.24\textwidth}
  \caption{Italy}
  \label{fig:IT_Supply}

\includegraphics[width=\maxwidth]{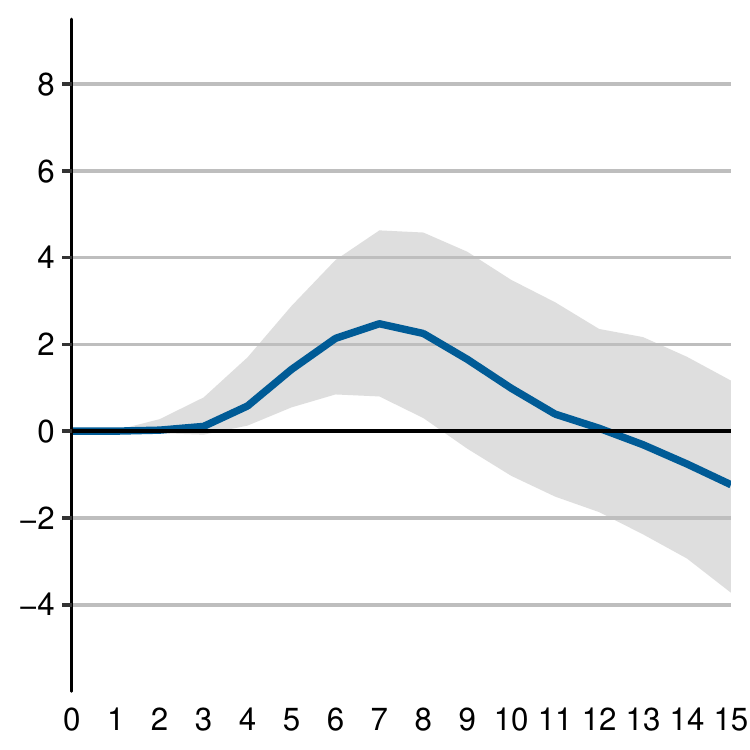} 

  \end{subfigure}
  \begin{subfigure}{.24\textwidth}
  \caption{Portugal}
  \label{fig:PT_Supply}

\includegraphics[width=\maxwidth]{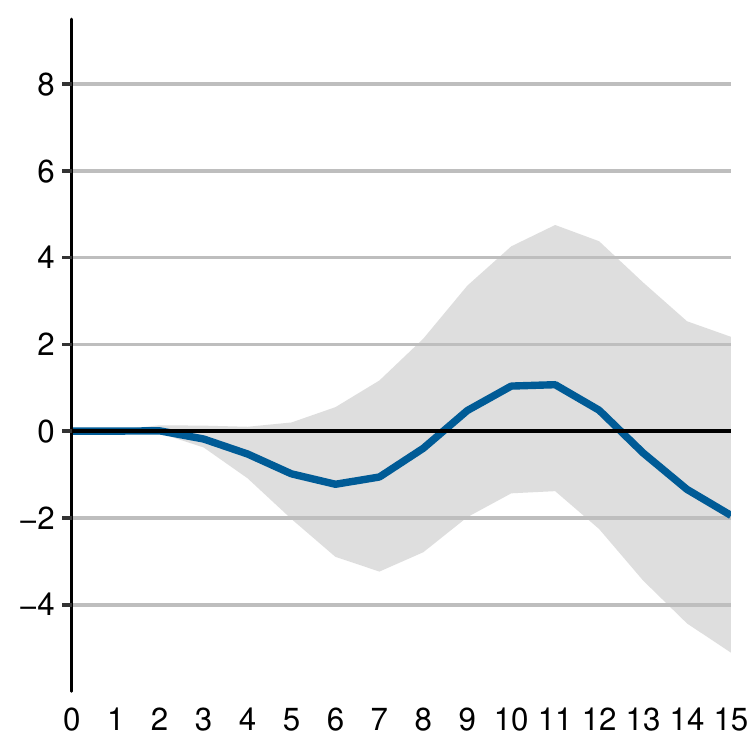} 

  \end{subfigure}
  \begin{subfigure}{.24\textwidth}
  \caption{Spain}
  \label{fig:ES_Supply}

\includegraphics[width=\maxwidth]{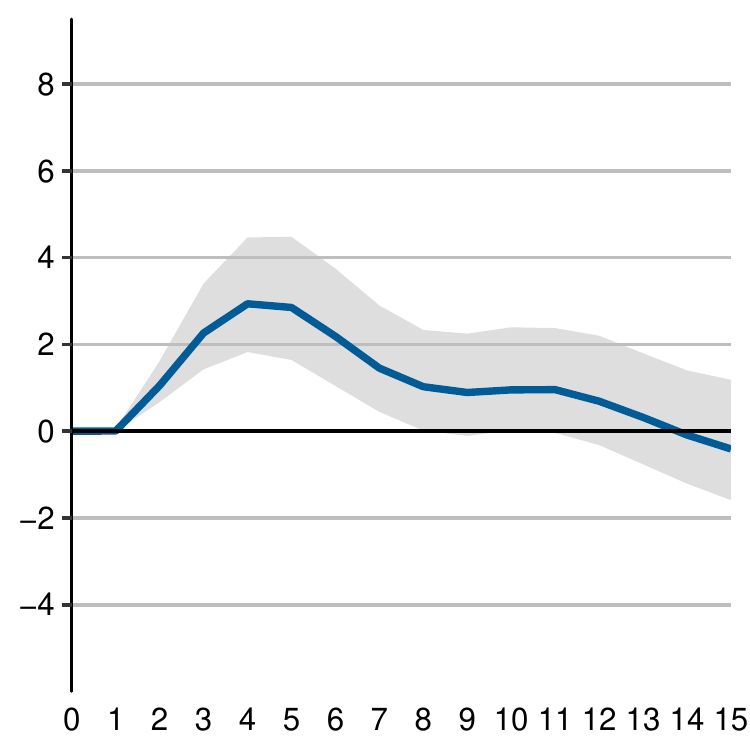} 

  \end{subfigure}
\caption{Impulse Responses of Loan Supply to an EONIA Shock. \textit{Note}: The chart plots the responses of loans supply to a expansionary 25 basis point monetary policy shock in the form of a decrease of the EONIA rate. The responses are divided by the shock itself and thus normalized, which yields a percentage point representation on the y-axis. The \textit{solid blue line} represents the median response along the \textit{grey shaded area} which covers the Bayesian credible interval between the 16th and 84th quantile. The x-axis steps are monthly.}
\label{fig:irf_Supply}
\end{figure}

%LS
Figure~\ref{fig:irf_Supply} shows the responses of loan supply. In theory, a negative shift in the policy variable should lead to more loan supply. Indeed, the general picture shows positive responses to the shock. Thus, one could infer that the supply conditions are loosened. However, a very heterogeneous picture between the countries arises. Austria and Spain show a instantaneous and strong effect, where the median peaks around 6\% and 3\% increases in credit supply, respectively. Both displaying long effects up to 14 months. Germany, Greece and Italy also exhibit also a positive effect between 2\%-4\%, but the response is delayed by about 5-6 months and dies out at around 13 months. The delayed effect can be explained by the cyclical behavior of credit supply, which could be stronger in the mentioned countries \parencite{giannone:2009}. In the case of Belgium, there is limited evidence regarding a positive effect on loan supply. Portugal displays no significant response at all. Looking at the supranational level, there is no clear center-periphery relationship displayed in figure~\ref{fig:irf_Supply}. It thus looks like Germany, Greece and Italy have been more apprehensive regarding their lending conditions.

%Loan Supply

\begin{figure}[h!]
  \begin{subfigure}{.12\textwidth}

  \caption*{}
  \label{fig:dummy5}
  \end{subfigure}
  \begin{subfigure}{.24\textwidth}
  \caption{Austria}
  \label{fig:AT_demand}

\includegraphics[width=\maxwidth]{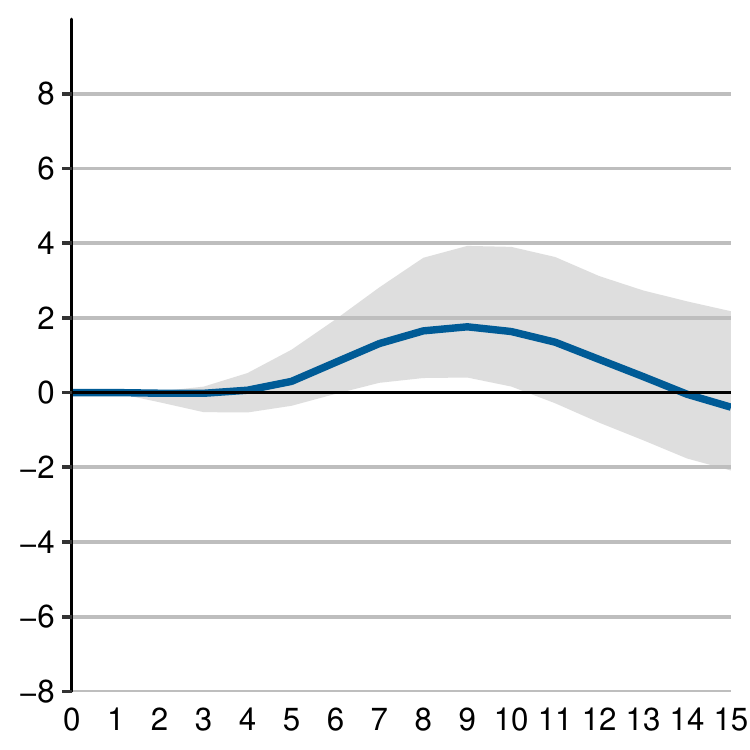} 

  \end{subfigure}
  \begin{subfigure}{.24\textwidth}
  \caption{Belgium}
  \label{fig:BE_demand}

\includegraphics[width=\maxwidth]{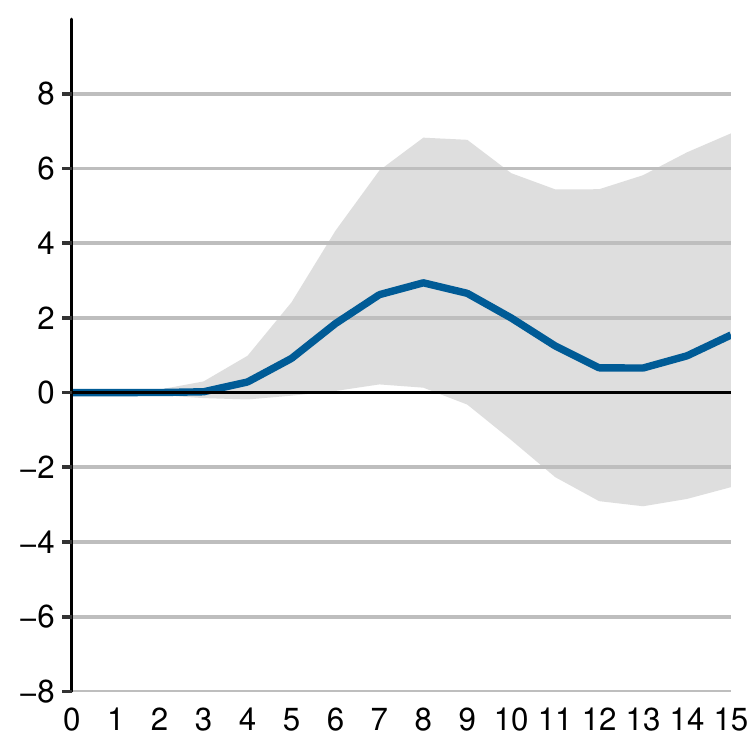} 

  \end{subfigure}
  \begin{subfigure}{.24\textwidth}
  \caption{Germany}
  \label{fig:DE_demand}

\includegraphics[width=\maxwidth]{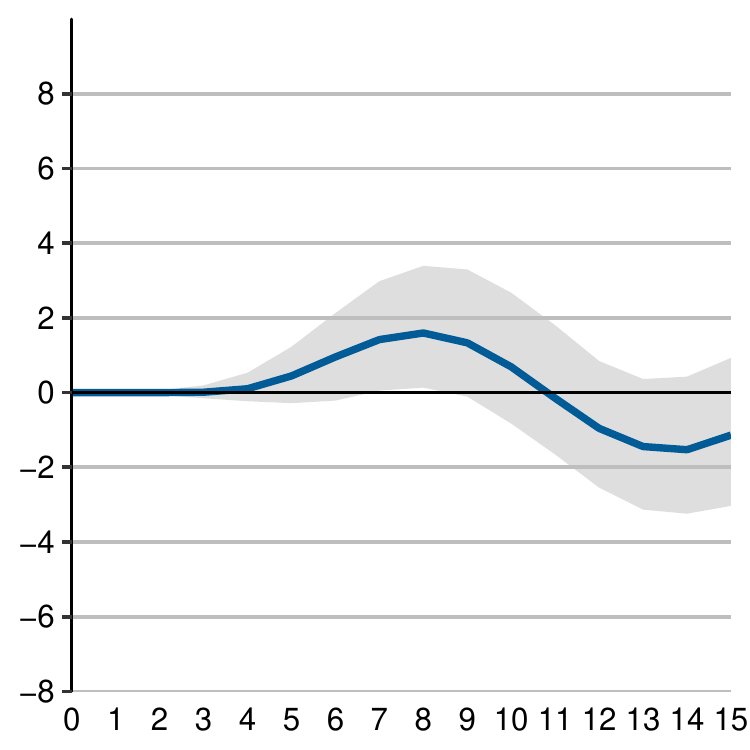} 

  \end{subfigure}
  \begin{subfigure}{.12\textwidth}

  \caption*{}
  \label{fig:dummy6}
  \end{subfigure}
  
  \vspace{0.3cm}
  
  \begin{subfigure}{.24\textwidth}
  \caption{Greece}
  \label{fig:GR_demand}

\includegraphics[width=\maxwidth]{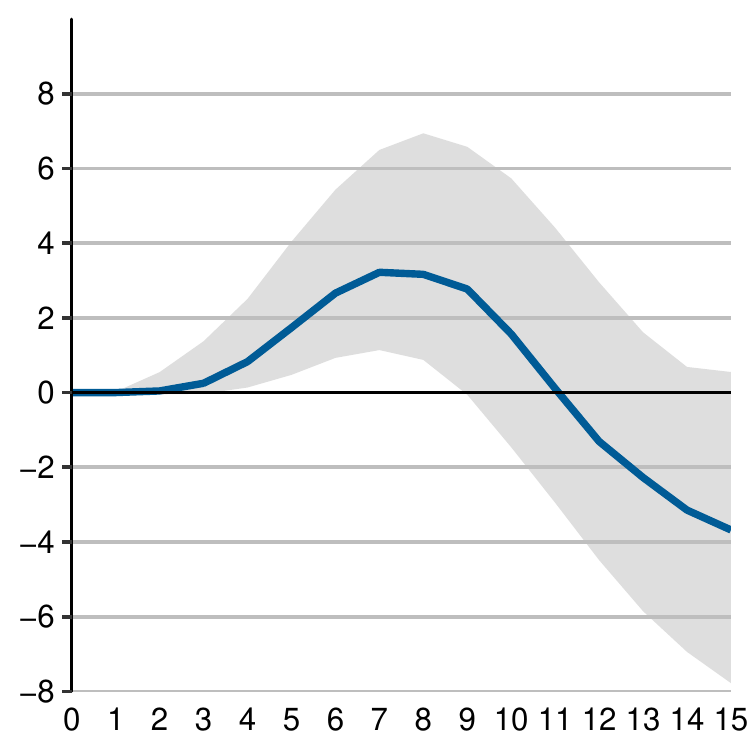} 

  \end{subfigure}
    \begin{subfigure}{.24\textwidth}
  \caption{Italy}
  \label{fig:IT_demand}

\includegraphics[width=\maxwidth]{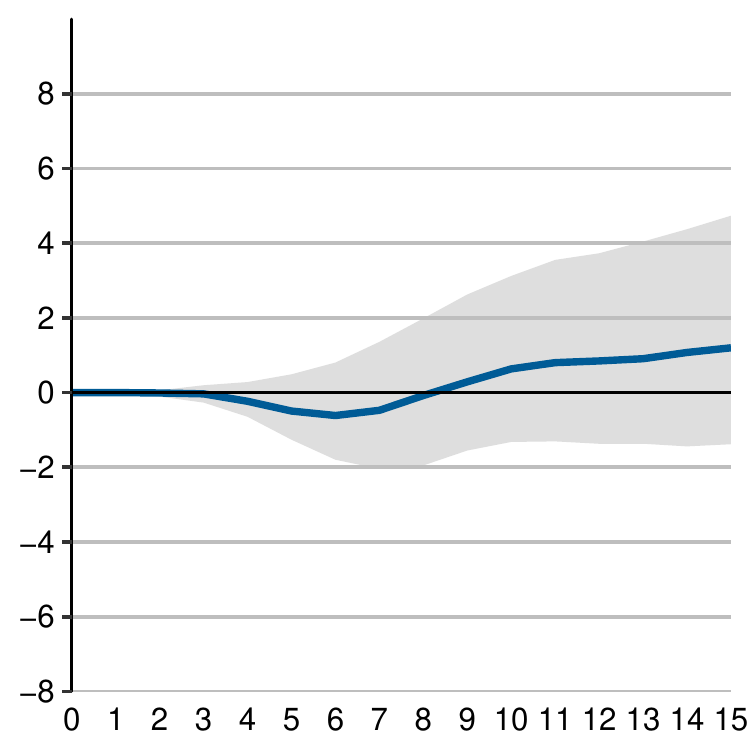} 

  \end{subfigure}
  \begin{subfigure}{.24\textwidth}
  \caption{Portugal}
  \label{fig:PT_demand}

\includegraphics[width=\maxwidth]{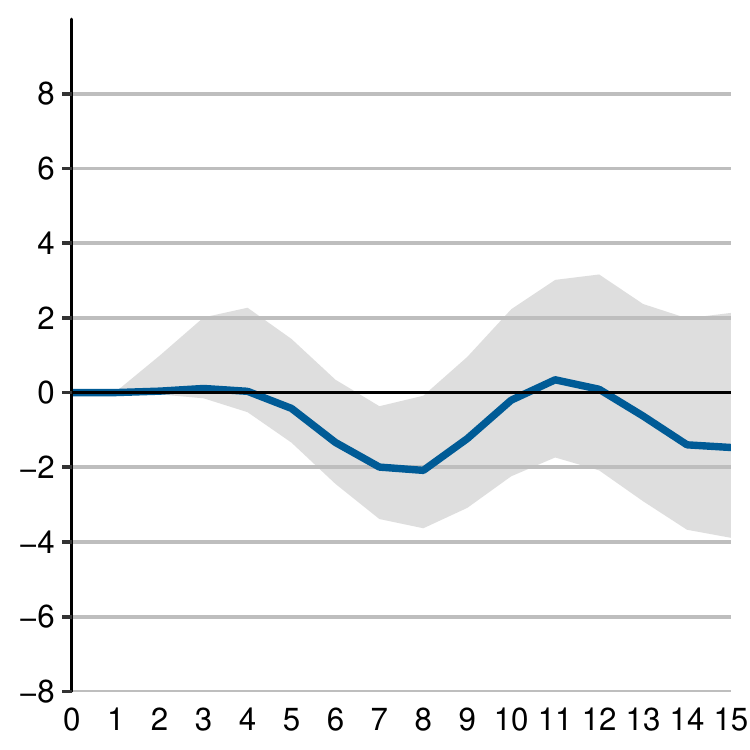} 

  \end{subfigure}
  \begin{subfigure}{.24\textwidth}
  \caption{Spain}
  \label{fig:ES_demand}

\includegraphics[width=\maxwidth]{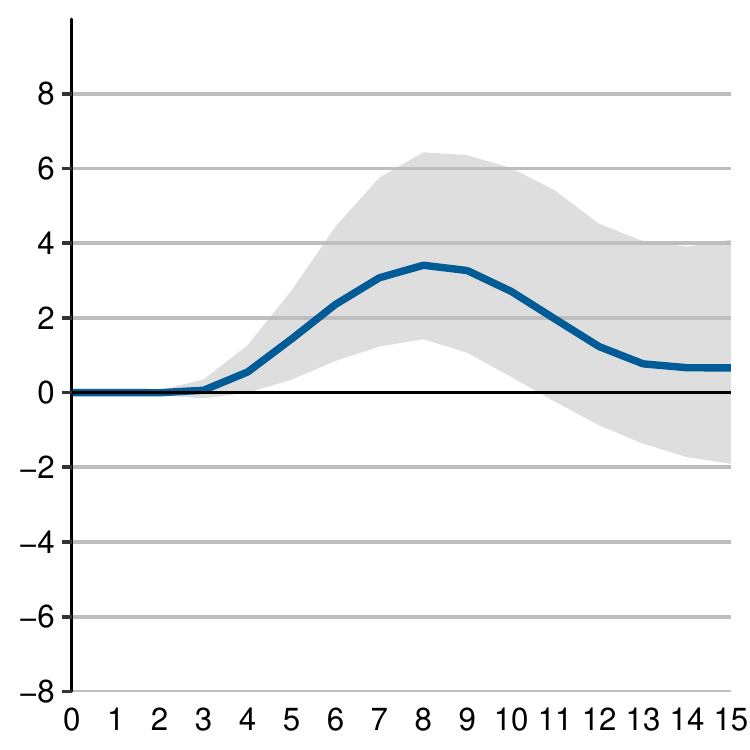} 

  \end{subfigure}
\caption{Impulse Responses of Loan Demand to an EONIA Shock. \textit{Note}: The chart plots the responses of loans demand to a expansionary 25 basis point monetary policy shock in the form of a decrease of the EONIA rate. The responses are divided by the shock itself and thus normalized, which yields a percentage point representation on the y-axis. The \textit{solid blue line} represents the median response along the \textit{grey shaded area} which covers the Bayesian credible interval between the 16th and 84th quantile. The x-axis steps are monthly.}
\label{fig:irf_demand}
\end{figure}

%LD
Figure~\ref{fig:irf_demand} displays the findings on loan demand reactions with respect to an expansionary monetary policy shock. Similar to loan supply, the theory predicts a positive response to a negative shock. The overall effects are small, but still positive. In fact, Austria, Belgium, Germany, Greece and Spain show a similar pattern of a positive response with around 5 to 7 months delay. The median response peaks at 2\% to 3\% credit demand increase and dies out between 11 to 14 months. Such a behavior is also mentioned by \textcite{praet:2017}, who explains that investments indeed react slowly to low interest rates, but that the ECB's monetary policy is helping in boosting investment. As most investments are financed by loans, the demand for it too reacts slowly to low interest rates. The outliers are Portugal and Italy with no significant reaction at all. As with loan supply, there is no clear north-south-disparity between the countries.

%Loan Demand

\begin{figure}[h!]
  \begin{subfigure}{.12\textwidth}

  \caption*{}
  \label{fig:dummy1}
  \end{subfigure}
  \begin{subfigure}{.24\textwidth}
  \caption{Austria}
  \label{fig:AT_GDP}

\includegraphics[width=\maxwidth]{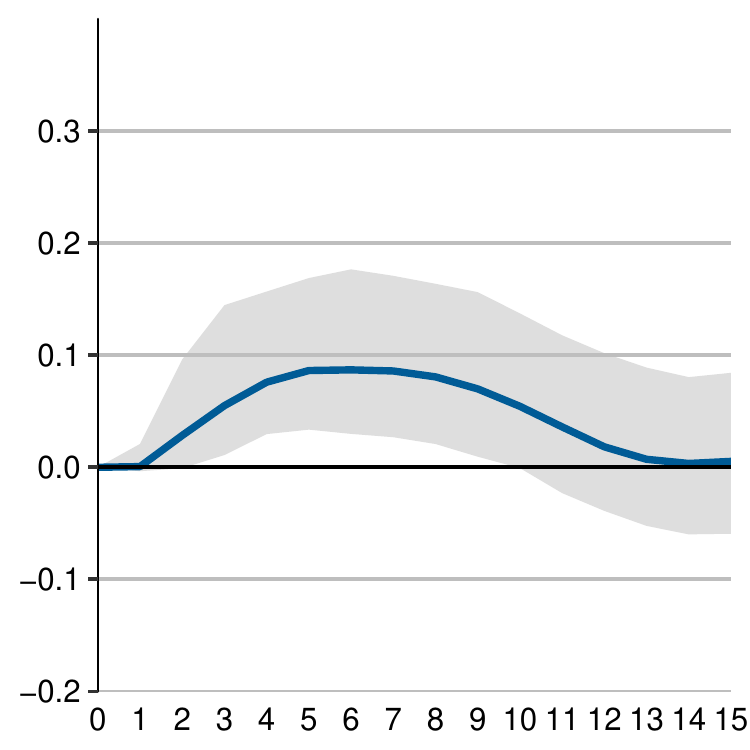} 

  \end{subfigure}
  \begin{subfigure}{.24\textwidth}
  \caption{Belgium}
  \label{fig:BE_GDP}

\includegraphics[width=\maxwidth]{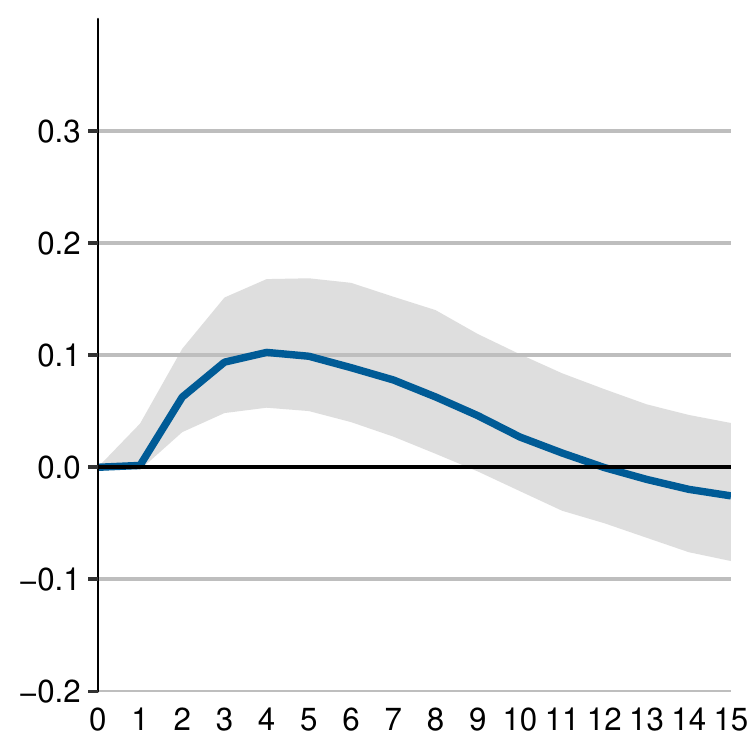} 

  \end{subfigure}
  \begin{subfigure}{.24\textwidth}
  \caption{Germany}
  \label{fig:DE_GDP}

\includegraphics[width=\maxwidth]{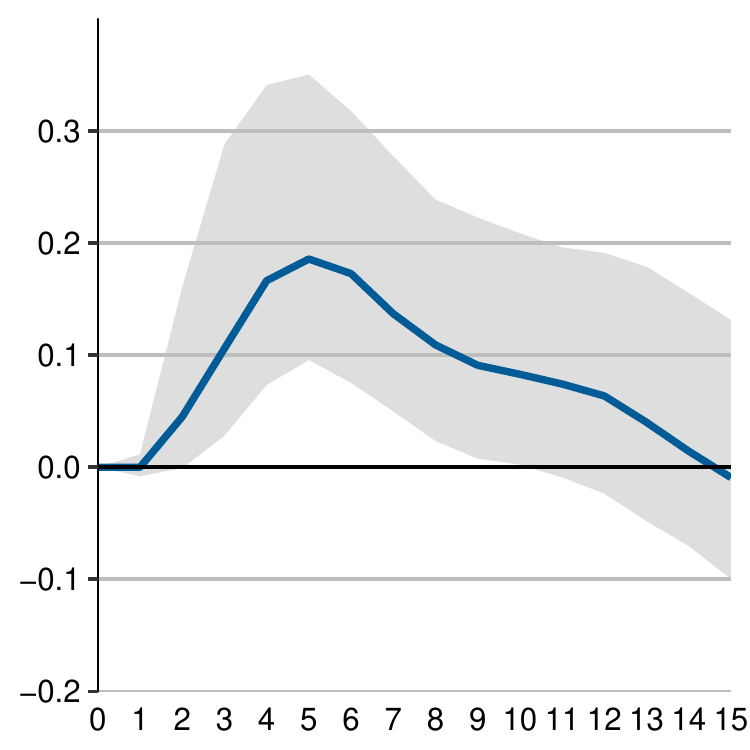} 

  \end{subfigure}
  \begin{subfigure}{.12\textwidth}

  \caption*{}
  \label{fig:dummy2}
  \end{subfigure}
  
  \vspace{0.3cm}
  
  \begin{subfigure}{.24\textwidth}
  \caption{Greece}
  \label{fig:GR_GDP}

\includegraphics[width=\maxwidth]{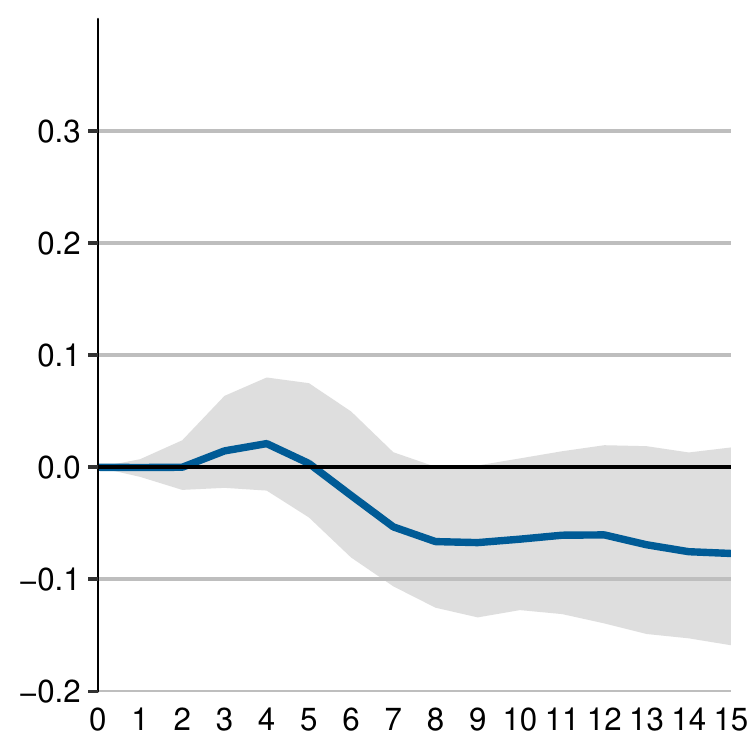} 

  \end{subfigure}
    \begin{subfigure}{.24\textwidth}
  \caption{Italy}
  \label{fig:IT_GDP}

\includegraphics[width=\maxwidth]{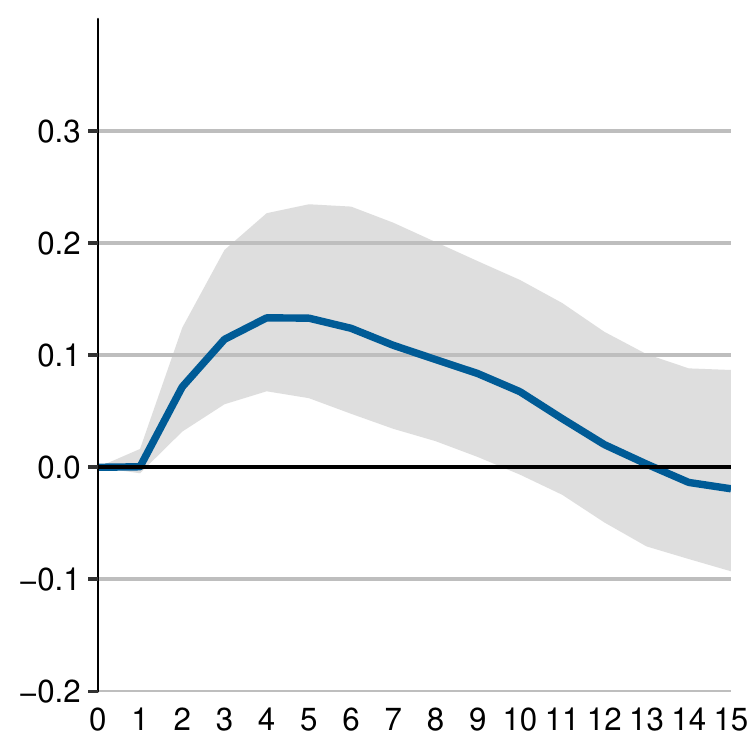} 

  \end{subfigure}
  \begin{subfigure}{.24\textwidth}
  \caption{Portugal}
  \label{fig:PT_GDP}

\includegraphics[width=\maxwidth]{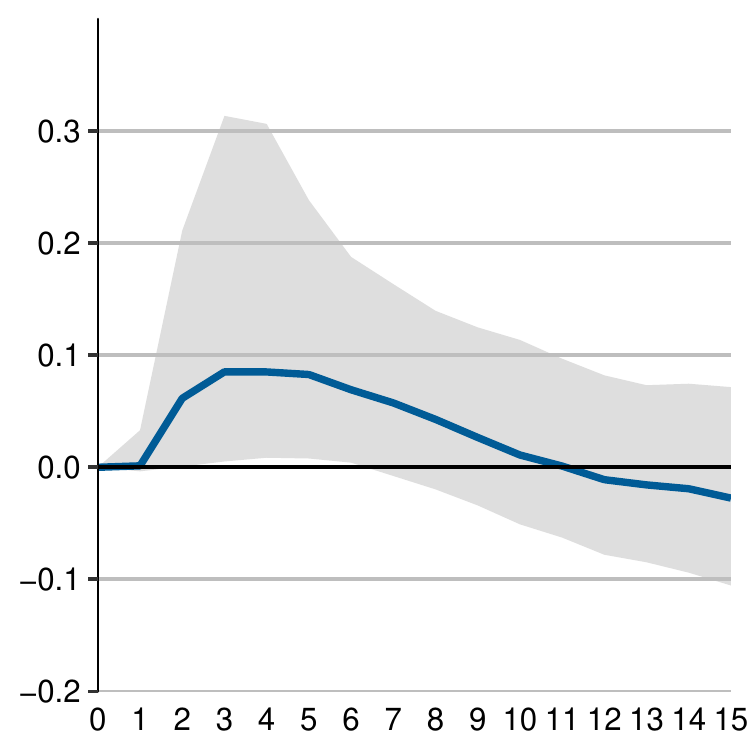} 

  \end{subfigure}
  \begin{subfigure}{.24\textwidth}
  \caption{Spain}
  \label{fig:ES_GDP}

\includegraphics[width=\maxwidth]{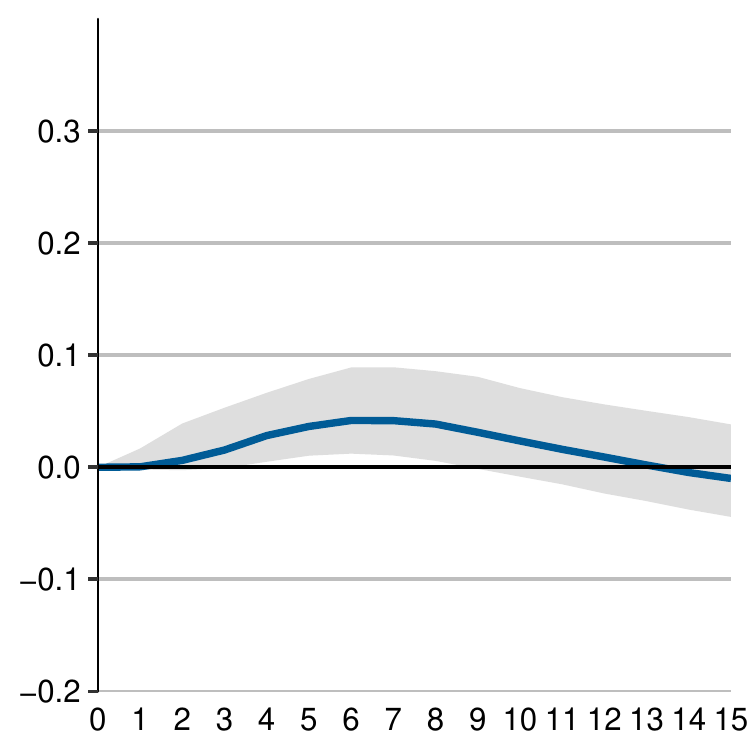} 

  \end{subfigure}
\caption{Impulse Responses of GDP Growth to an EONIA Shock. \textit{Note}: The chart plots the responses of GDP growth to a expansionary 25 basis point monetary policy shock in the form of a decrease of the EONIA rate. The responses are divided by the shock itself and thus normalized, which yields a percentage point representation on the y-axis. The \textit{solid blue line} represents the median response along the \textit{grey shaded area} which covers the Bayesian credible interval between the 16th and 84th quantile. The x-axis steps are monthly.}
\label{fig:irf_gdp}
\end{figure}

%GDP
The third result is displayed in figure~\ref{fig:irf_gdp} and reports the impact on GDP growth. The overall picture is inline with the theory of the broad credit channel and with the literature \parencite{boeckx:2017, ciccarelli:2013}. Increased credit support policies elevate GDP growth after a few months delay. In particular, Austria, Belgium, Germany and Italy show a significant increase with the median response peaking between 0.1\% and 0.2\% and the effect dying out at around 12 to 15 month. Spain and Portugal display only limited evidence on a positive effect, whereas Greece's response of GDP growth is not significant at all. The findings not only show a working bank lending channel, but also a north-south-disparity. This could be due to additional investment incentives and a general stronger economy in the north. This assumption is inline with \textcite{boeckx:2017} findings that retail banks in the periphery are constrained by their financial fragility. Even though Italy cannot be considered a financially sound country, the degree of bank market concentration could be the driver of the positive effect of output. The large number of small mutual savings banks amplify the effect of the bank lending channel as those intermediaries are more prone to changes in their costs of funding \parencite{desantis:2013}. More recently, \textcite{burriel:2018} assert these findings on the north-south-disparity by concluding that the effect of an unconventional monetary policy shock on output depends on the resilience of the banking sector.\footnote{All other results - HICP, CISS, government bond yields, MFI deposits, MRO and Euro Stoxx - can be found in the appendix, figure~\ref{fig:irf_HICP} - \ref{fig:irf_EA}.}

%GDP Growth

\subsection{Robustness Checks}
\label{sec:robust}

\begin{figure}[H]
  \begin{subfigure}{.32\textwidth}
  \caption{Loan Demand}
  \label{fig:rc_ld}

\includegraphics[width=\maxwidth]{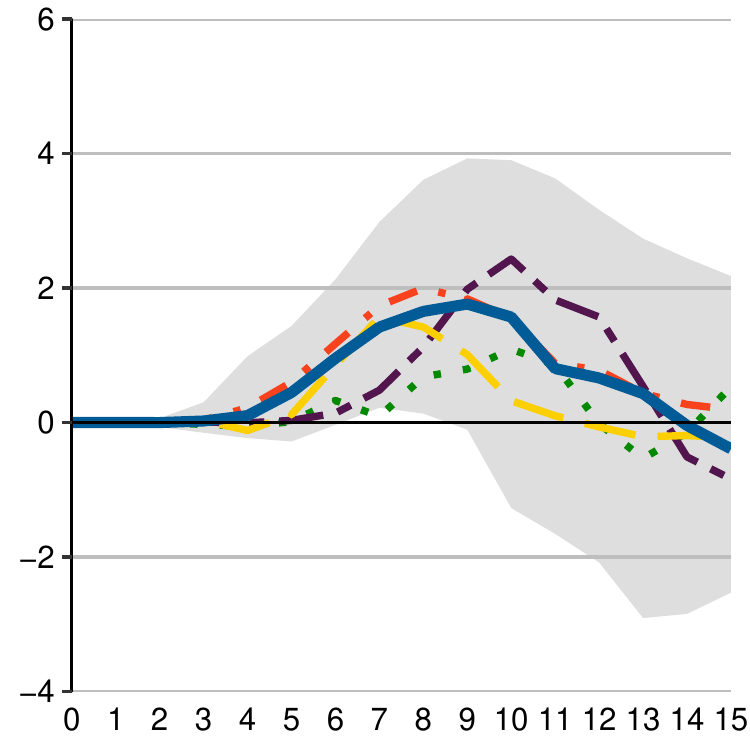} 

  \end{subfigure}
  \begin{subfigure}{.32\textwidth}
  \caption{Loan Supply}
  \label{fig:rc_ls}

\includegraphics[width=\maxwidth]{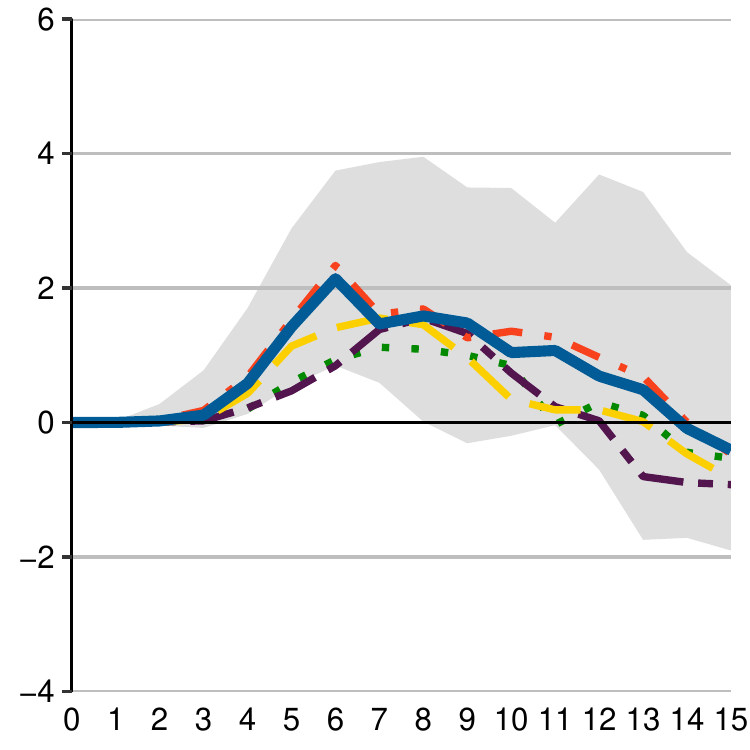} 

  \end{subfigure}
  \begin{subfigure}{.32\textwidth}
  \caption{GDP Growth}
  \label{fig:rc_gdp}

\includegraphics[width=\maxwidth]{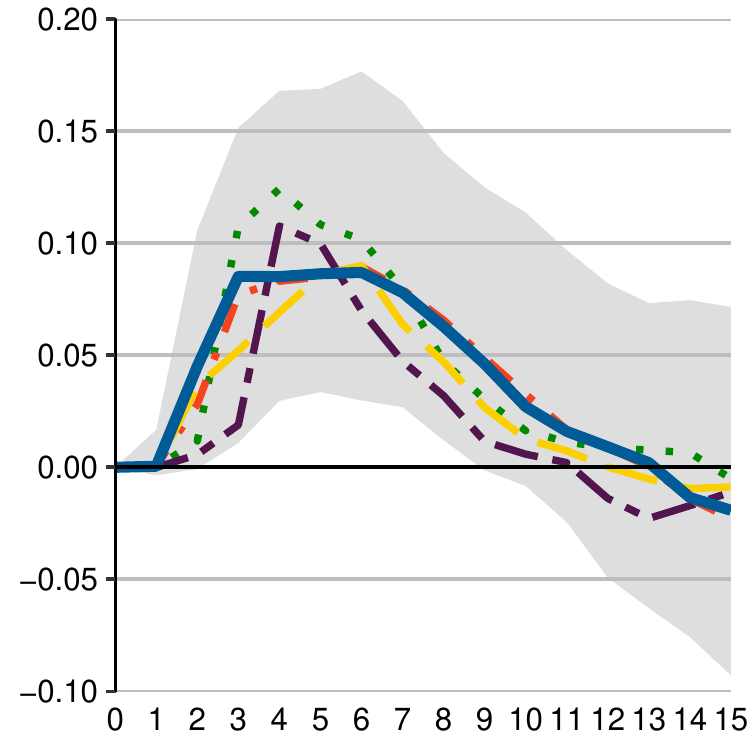} 

  \end{subfigure}
  
  \vspace{0.3cm}
  
  \begin{subfigure}{.32\textwidth}
  \caption{HICP}
  \label{fig:rc_hicp}

\includegraphics[width=\maxwidth]{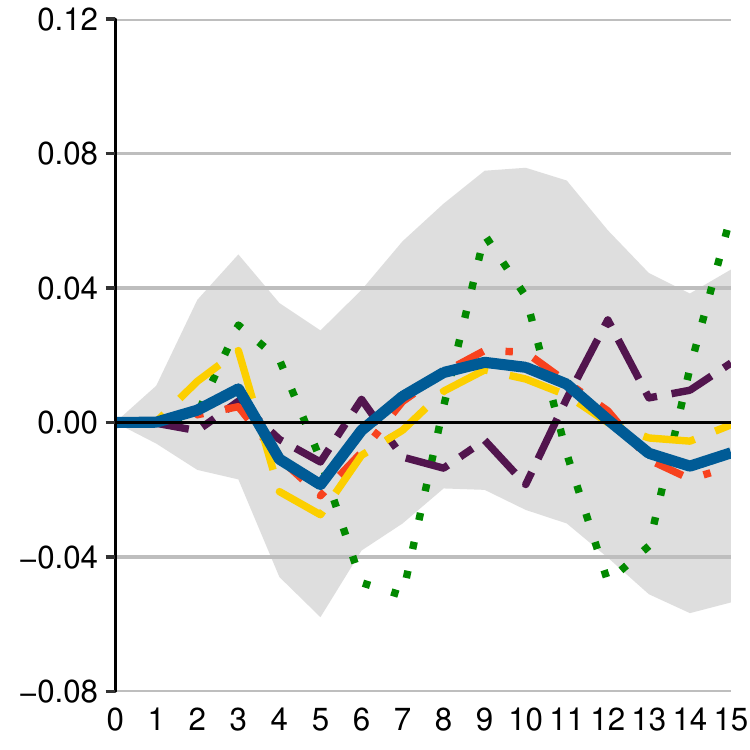} 

  \end{subfigure}
  \begin{subfigure}{.32\textwidth}
  \caption{CISS}
  \label{fig:rc_ciss}

\includegraphics[width=\maxwidth]{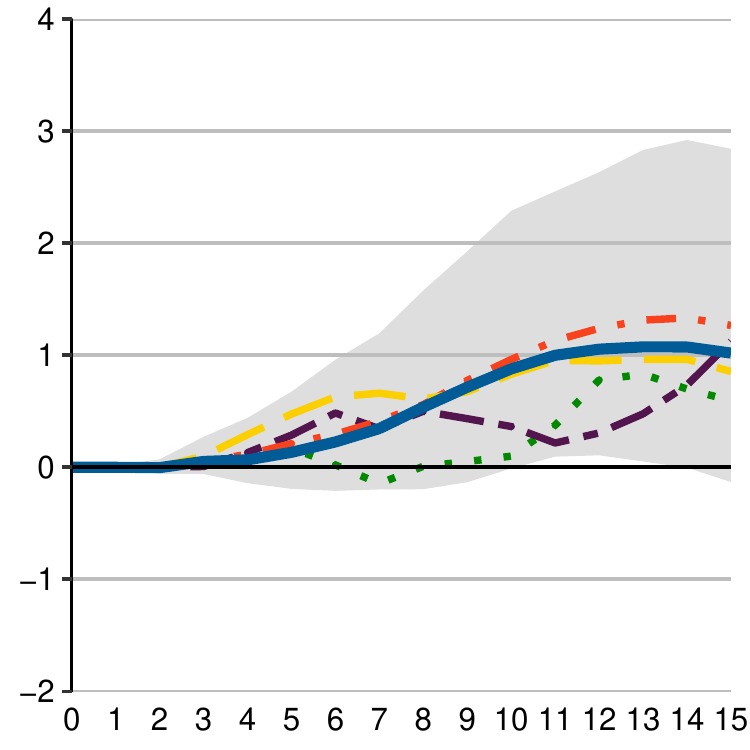} 

  \end{subfigure}
  \begin{subfigure}{.32\textwidth}
  \caption{Gov. Bond Yields}
  \label{fig:rc_gby}

\includegraphics[width=\maxwidth]{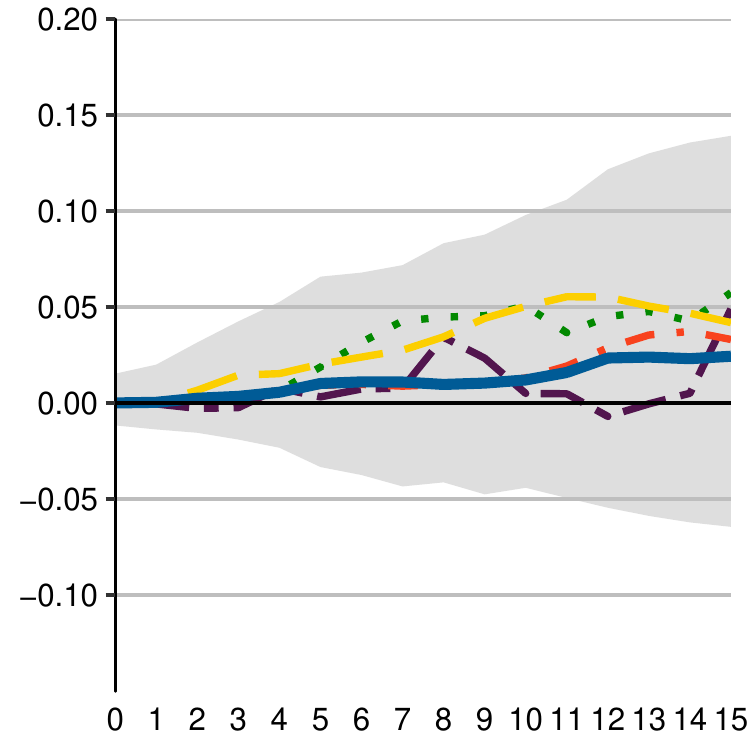} 

  \end{subfigure}
    
  \vspace{0.3cm}
  
  \begin{subfigure}{.32\textwidth}
  \caption{MFI Deposits}
  \label{fig:rc_deposits}

\includegraphics[width=\maxwidth]{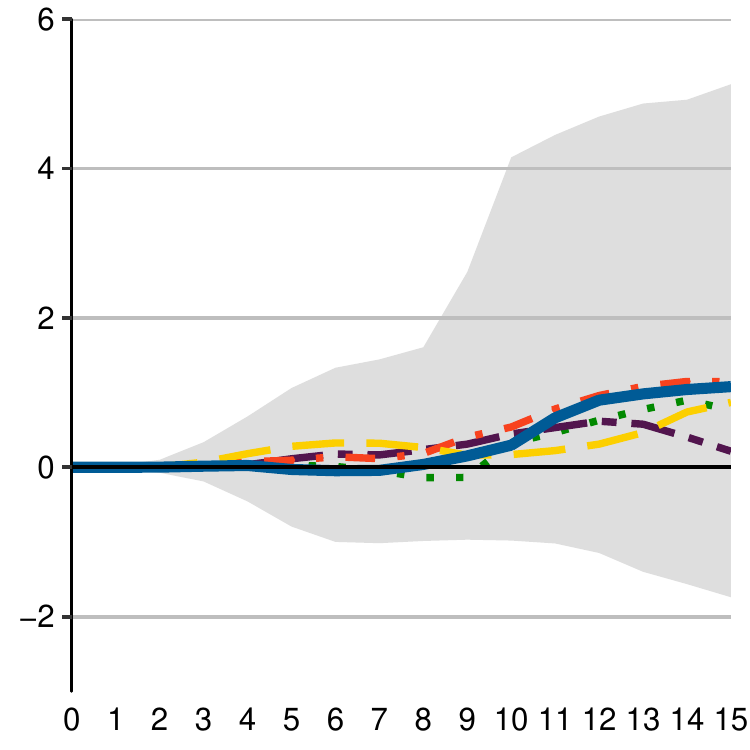} 

  \end{subfigure}
  \begin{subfigure}{.32\textwidth}
  \caption{MRO rate}
  \label{fig:rc_mro}

\includegraphics[width=\maxwidth]{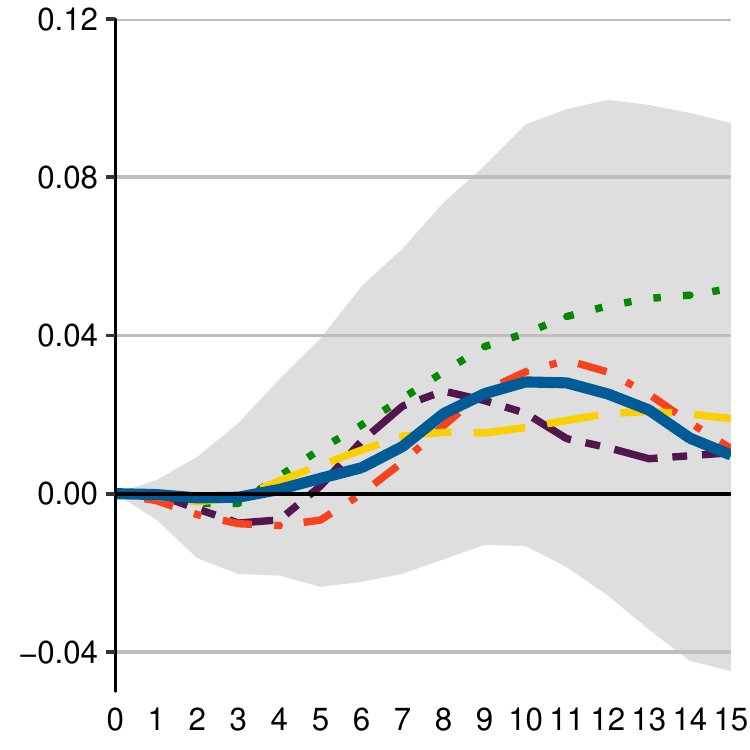} 

  \end{subfigure}
  \begin{subfigure}{.32\textwidth}
  \caption{Euro Stoxx}
  \label{fig:rc_stoxx}

\includegraphics[width=\maxwidth]{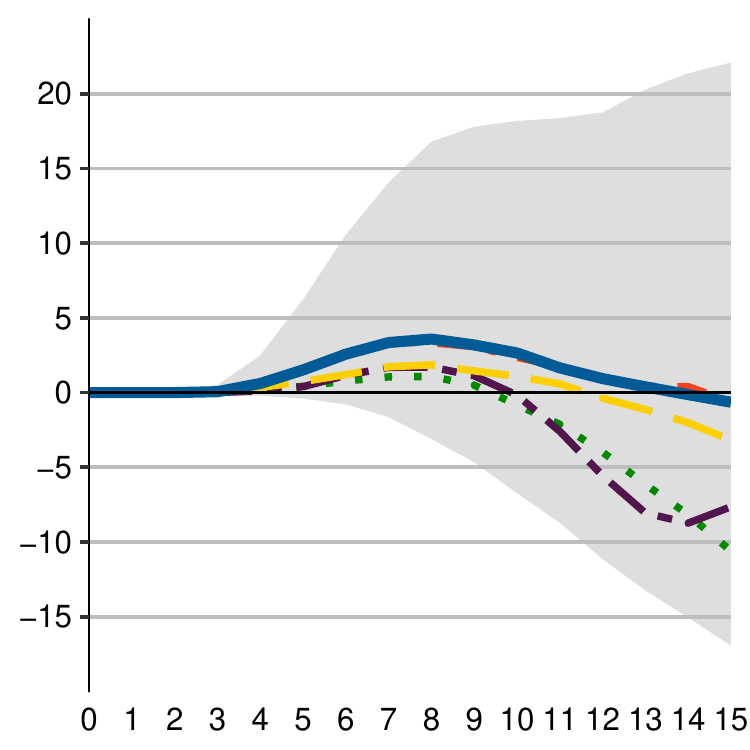} 

  \end{subfigure}
  
  \vspace{0.3cm}
  
  \begin{subfigure}{.20\textwidth}

  \caption*{}
  \end{subfigure}
  \begin{subfigure}{.70\textwidth}
  \includegraphics[width=60mm, scale=0.9]{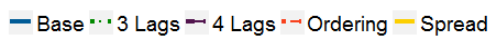}
  \end{subfigure}
  \begin{subfigure}{.10\textwidth}

  \caption*{}
  \end{subfigure}
  \vspace{-0.8cm}
\caption{Robustness Check Impulse Responses. \textit{Note}: The chart plots the median responses of all countries for all available variables to a expansionary 25 basis point monetary policy shock. The responses are divided by the shock itself and thus normalized. The \textit{solid blue line} represents the median response along the \textit{grey shaded area} which covers the Bayesian credible interval between the 16th and 84th quantile for the base model. The \textit{green} and \textit{purple line} are the base models with 3 and 4 lags respectively. The \textit{orange dot-dash line} is based on a changed Cholesky ordering. The \textit{yellow line with the long dashes} uses the EONIA-MRO spread as shock variable. The x-axis steps are monthly.}
\label{fig:irf_rc}
\end{figure}

Figure~\ref{fig:irf_rc} displays the median impulse responses of all variables for a set of alternative specifications alongside the credible intervals of the base model. Thus, the reader can compare the performance and its statistical significance.\footnote{The idea of median impulse responses is borrowed from \textcite{huber:2015}.}

The first two tests change the number of chosen lags to three and four respectively. The model with three lags yields loan effects which would have been a bit lower compared to the base model, however, the GDP growth response would have been more pronounced. Nearly the same holds for the model with four lags. The effect on the responses with different lags can clearly be seen when comparing the response of loan demand from the base model to the four lag model. The effect is shifted to the right. However, this paper uses two lags as a standard choice since the other two models are less stable\footnote{Multiple eigenvalues are above one in both models.} and as the number of observations is relatively small, the model with four lags pushes the framework to its feasible limits.

As the model with two lags is the most stable one, all further robustness checks will be carried out with the baseline model. The third robustness check changes the variable ordering of the Cholesky decomposition. Thus, the assumptions around the second pillar of monetary policy strategy will be loosened up a bit. In other words, the loan variables will now be ordered below the shock variable. This would indicate that the ECB still observes prices, output and the stress in financial markets before conducting further policy decisions, however, these decisions will not be based on the outcome of the Bank Lending Survey. Indeed, the loan variables are now defined to move contemporaneously with the monetary policy shock. However, the findings are virtually the same.

%Robustness Figure

In the final test the EONIA rate will be exchanged for the spread between the MRO and the EONIA rate. The ECB increased the demand and supply for liquidity with the deployed fixed-rate full allotment, which in turn exerted a downward pressure on the EONIA rate. The ongoing provision of money led to excess liquidity, which further opened the spread between the main policy rate and the overnight index. These effects can be especially observed during the 3-year longer-term refinancing operations \parencite{albertazzi:2016, boeckx:2017, lenza:2010}. Thus, the spread should be a 'pure' indicator of credit enhancing policies. Re-estimating the model with this shock variable yields nearly the same results as in the base model, with responses of the loan variables and GDP growth being a bit lower.

\section{Conclusion}
\label{sec:con}

With the tails of the financial crisis still being noticeable, this paper analysis the bank lending channel via the effects of an unconventional monetary policy shock. The sample period covers the start of the crisis till the latest available time point and thus includes all policy measures taken by the ECB. However, this paper focuses on the credit enhancing policies which had a direct effect on the available liquidity to banks. The findings show that the enhanced credit support policies undertaken by the ECB helped to circumvent the impaired transmission channels via new channels to stimulate bank lending and thus output.

The main task for the analysis of the channel is the identification of a suitable policy variable and the developments of loan demand and supply. The former is identified via the EONIA rate, as this rate has shown to posses the properties needed to reflect the credit enhancing policies taken by the ECB. The disentanglement of loan demand and supply is achieved via the usage of the Bank Lending Survey. This rich database delivers the needed variables for a stringent separation of demand and supply. 

The paper contributes to the literature via a separation of the usual aggregated responses into country-level responses and thus uncovering a north-south disparity between Austria, Belgium, Germany, Italy and Greece, Portugal, Spain. The results suggest that the bank lending channel indeed is operational throughout the sample period. The increased liquidity by the ECB helped foster output growth as a second round effect. Moreover, a clear indication of country-specific effects from an unified unconventional monetary policy shock can be observed. In a low interest environment, the positive responses of loan supply vary between 2\% to 4\% between the countries. With Germany, Greece and Italy being a bit more apprehensive regarding their lending conditions as the other countries. Same holds true for the responses of loan demand, with five out of seven countries displaying a positive effect between 2\% to 3\%. The findings for GDP growth are inline with the prediction of the bank lending channel. An expansionary monetary policy shock boosts output growth for several months. However, those results show a clear center-periphery-relationship, but the disparity can only be observed for the macroeconomic variable. This provides evidence that the banking sector and financial markets are highly interconnected and that banks can spread their risk across borders. However, macroeconomic policies are country-specific and thus center-periphery-relationships are easier to observe.

Implications from this paper can be drawn regarding theory and policy. In general, the findings corroborate the existing literature in highlighting the functionality of the bank lending channel during crisis periods. Furthermore confirming that stressed countries had a difficult time increasing the credit availability to the economy. This in turn would explain the hampered effects on GDP growth. In order to abolish such a north-south-disparity, fiscal policy is needed to create a situation in which companies are in a position where they can invest and thus boost economic activity. Summarizing, the one size fits all approach of the ECB helped all countries, in one way or another, to foster loan supply and partly output growth. However, unconventional monetary policy in a single-currency union with such heterogeneous countries also has its boundaries. This is the point were programs such as the "Europe 2020" strategy and the "Stability and Growth Pact" come into play. These endeavors will help to further strengthen the euro area economy via increased employment, production and thus output.

\newpage

\section*{Compliance with Ethical Standards}
\label{sec:comp}

\noindent Conflict of Interest: The author declares that he has no conflict of interest.\newline

\noindent Ethical approval: This article does not contain any studies with human participants or animals performed by the author.

\newpage

\printbibliography

\newpage

\begin{appendices}

% \counterwithin{table}{section}
% 
% \tocless\section{Tables}
% 
% <<tableminmax, eval=TRUE, echo=FALSE, results='asis'>>=
% latex(minmax_table[3:6], title="", file="", align="rrrr",
%       caption = "Descriptive Statistics for all Variables split up for all Countries",
%       header = colnames(minmax_table[3:6]),
%       rowname = rownames,
%       rgroup = sort(unique(minmax_table$ref_area)),
%       n.rgroup = no.rgroup,
%       ctable=FALSE,
%       longtable=TRUE,
%       lines.page=4000,
%       continued = "Table continued",
%       insert.bottom = "Source: ECB; Eurostat; OECD"
% )
% @
% 
% \newpage

% \counterwithin{figure}{section}
% 
% \tocless\section{Figures}

%HICP
\begin{figure}[h!]
  \begin{subfigure}{.12\textwidth}

  \caption*{}
  \label{fig:dummy7}
  \end{subfigure}
  \begin{subfigure}{.24\textwidth}
  \caption{Austria}
  \label{fig:AT_HICP}

\includegraphics[width=\maxwidth]{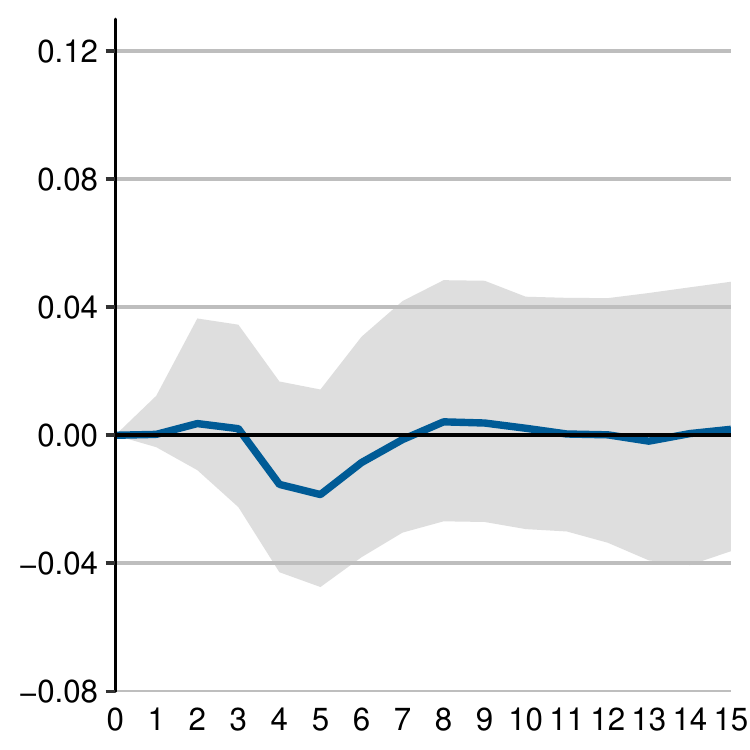} 

  \end{subfigure}
  \begin{subfigure}{.24\textwidth}
  \caption{Belgium}
  \label{fig:BE_HICP}

\includegraphics[width=\maxwidth]{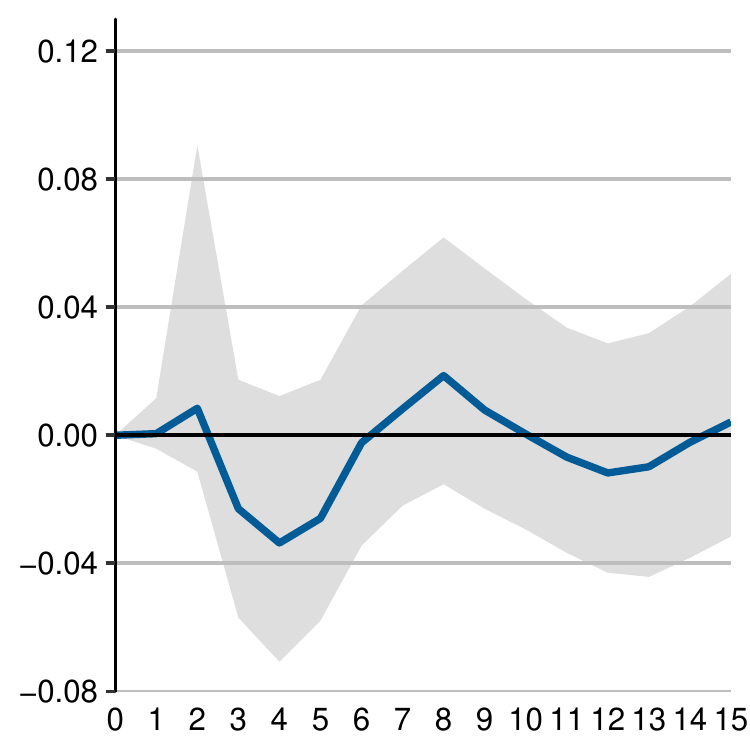} 

  \end{subfigure}
  \begin{subfigure}{.24\textwidth}
  \caption{Germany}
  \label{fig:DE_HICP}

\includegraphics[width=\maxwidth]{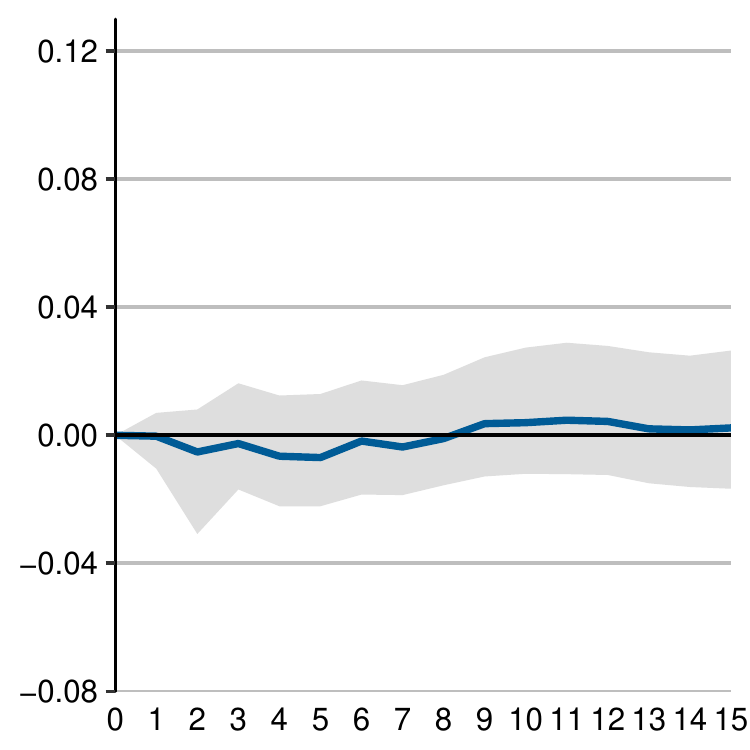} 

  \end{subfigure}
  \begin{subfigure}{.12\textwidth}

  \caption*{}
  \label{fig:dummy8}
  \end{subfigure}
  
  \vspace{0.5cm}
  
  \begin{subfigure}{.24\textwidth}
  \caption{Greece}
  \label{fig:GR_HICP}

\includegraphics[width=\maxwidth]{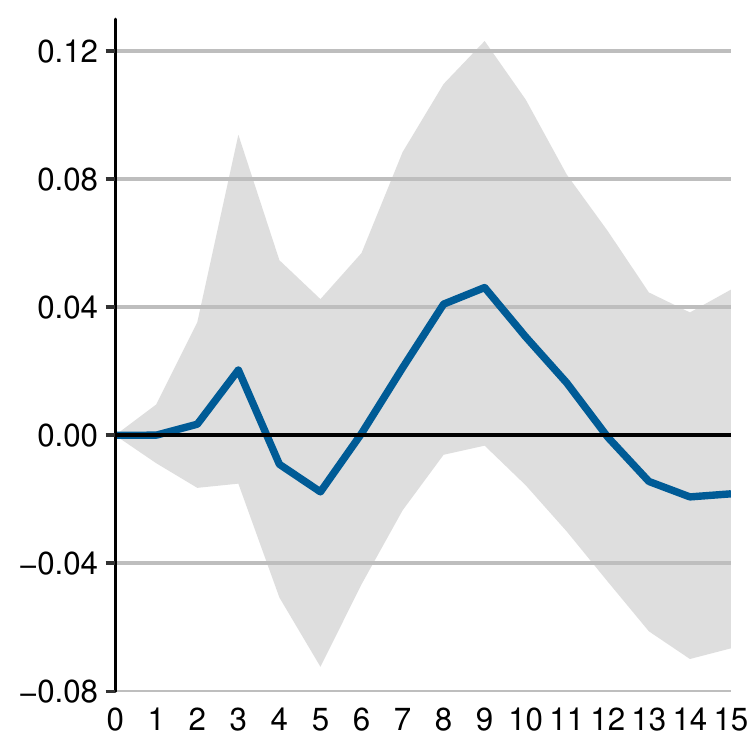} 

  \end{subfigure}
    \begin{subfigure}{.24\textwidth}
  \caption{Italy}
  \label{fig:IT_HICP}

\includegraphics[width=\maxwidth]{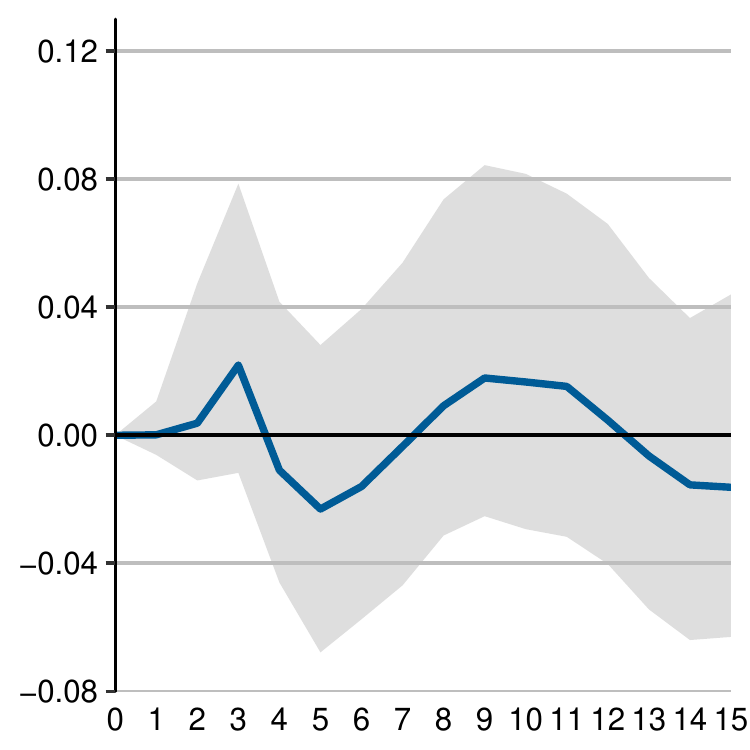} 

  \end{subfigure}
  \begin{subfigure}{.24\textwidth}
  \caption{Portugal}
  \label{fig:PT_HICP}

\includegraphics[width=\maxwidth]{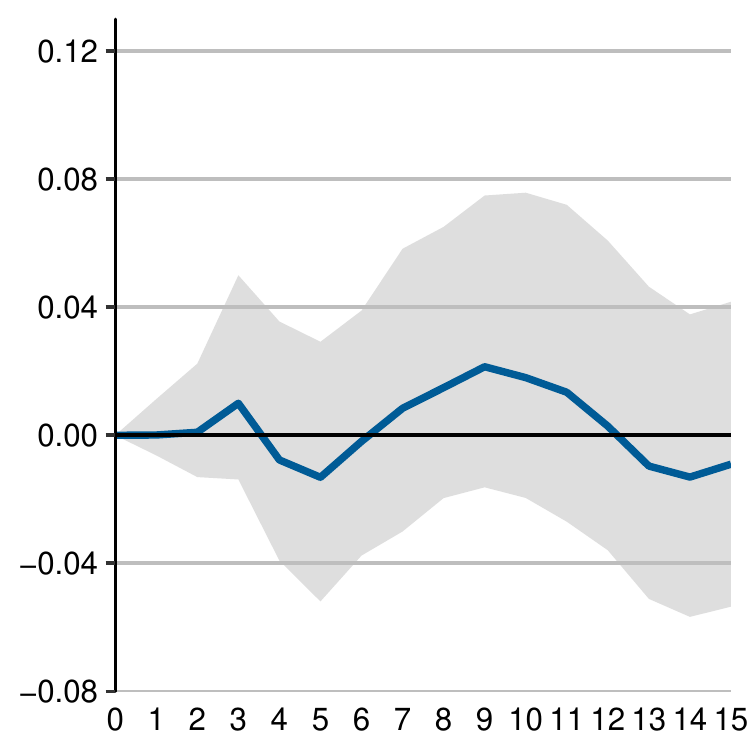} 

  \end{subfigure}
  \begin{subfigure}{.24\textwidth}
  \caption{Spain}
  \label{fig:ES_HICP}

\includegraphics[width=\maxwidth]{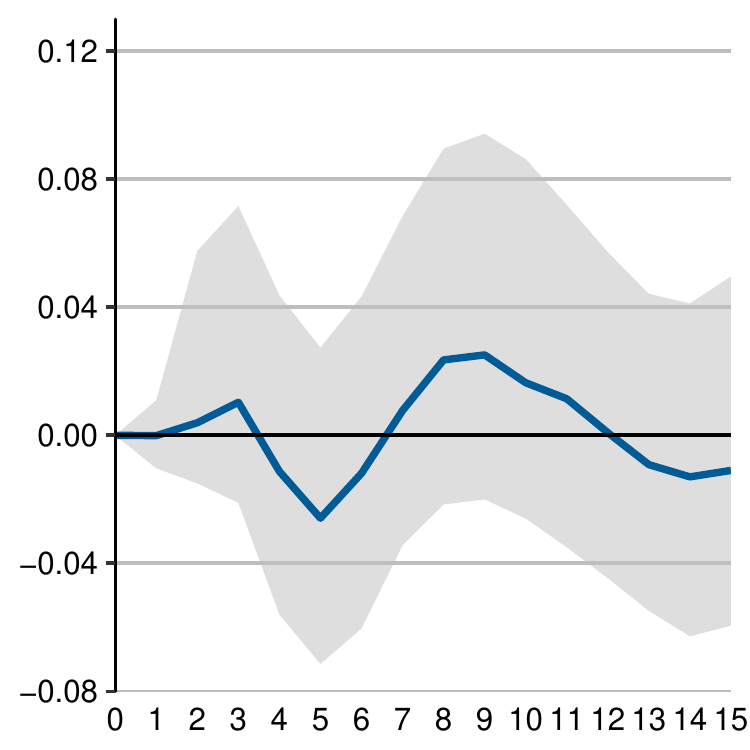} 

  \end{subfigure}
\caption{Impulse Responses of HICP to an EONIA Shock. \textit{Note}: The chart plots the responses of the Harmonised Index of Consumer Prices (HICP) to a expansionary 25 basis point monetary policy shock in the form of a decrease of the EONIA rate. The responses are divided by the shock itself and thus normalized, which yields a percentage point representation on the y-axis. The \textit{solid blue line} represents the median response along the \textit{grey shaded area} which covers the Bayesian credible interval between the 16th and 84th quantile. The x-axis steps are monthly.}
\label{fig:irf_HICP}
\end{figure}

%CISS
\begin{figure}[h!]
  \begin{subfigure}{.12\textwidth}

  \caption*{}
  \label{fig:dummy9}
  \end{subfigure}
  \begin{subfigure}{.24\textwidth}
  \caption{Austria}
  \label{fig:AT_CISS}

\includegraphics[width=\maxwidth]{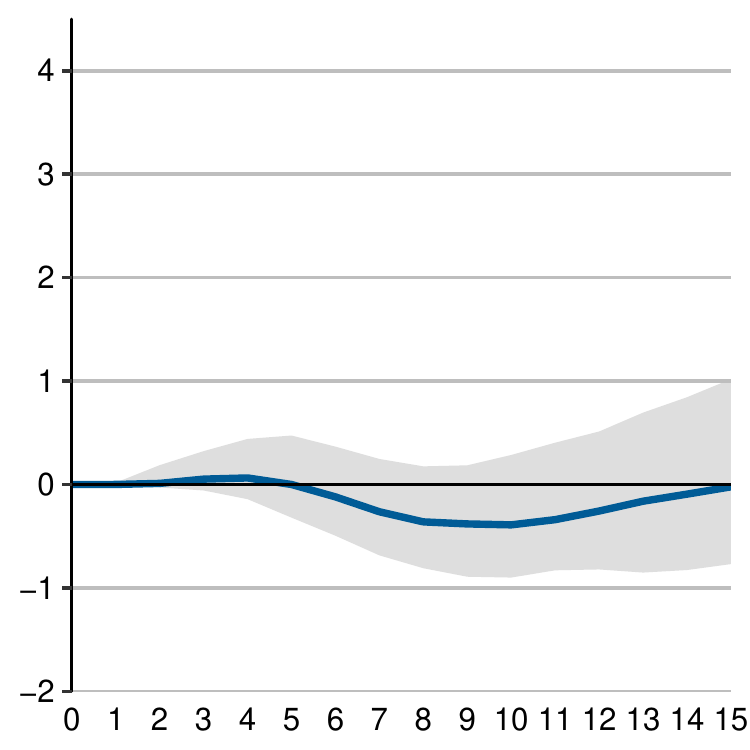} 

  \end{subfigure}
  \begin{subfigure}{.24\textwidth}
  \caption{Belgium}
  \label{fig:BE_CISS}

\includegraphics[width=\maxwidth]{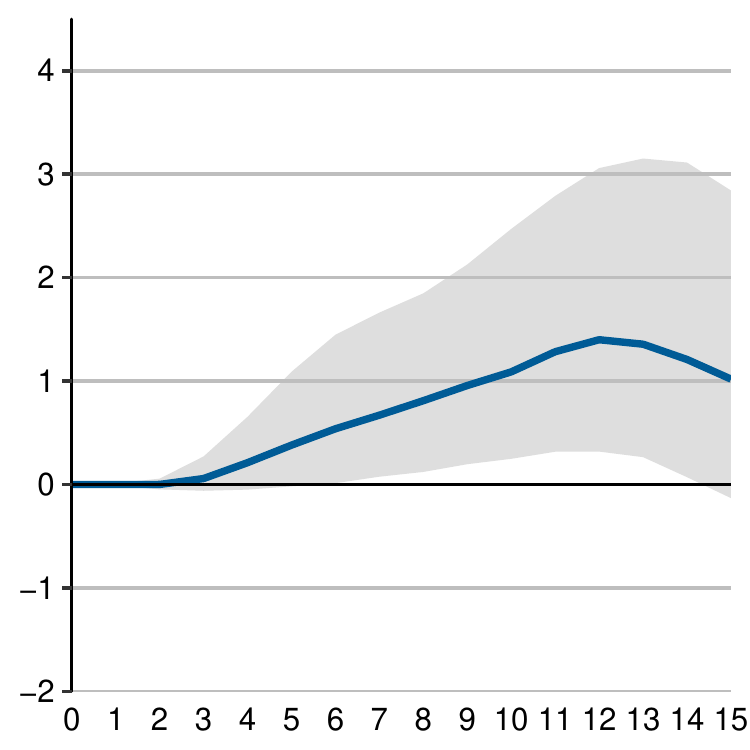} 

  \end{subfigure}
  \begin{subfigure}{.24\textwidth}
  \caption{Germany}
  \label{fig:DE_CISS}

\includegraphics[width=\maxwidth]{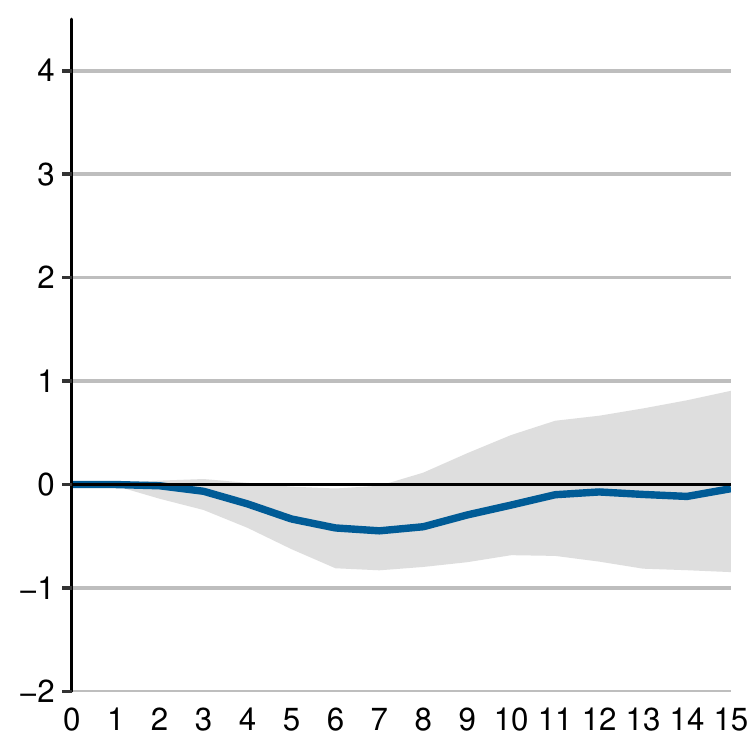} 

  \end{subfigure}
  \begin{subfigure}{.12\textwidth}

  \caption*{}
  \label{fig:dummy10}
  \end{subfigure}
  
  \vspace{0.5cm}
  
  \begin{subfigure}{.24\textwidth}
  \caption{Greece}
  \label{fig:GR_CISS}

\includegraphics[width=\maxwidth]{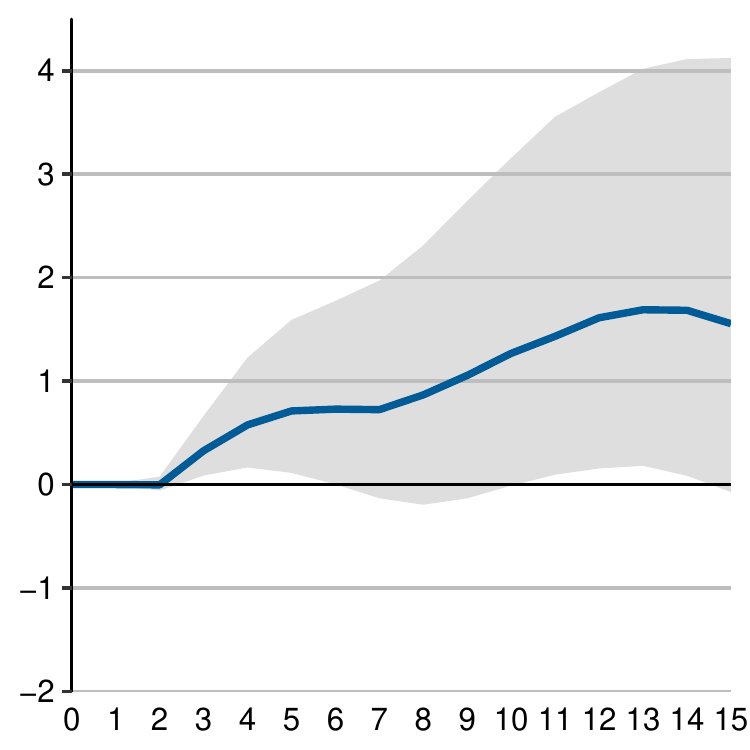} 

  \end{subfigure}
    \begin{subfigure}{.24\textwidth}
  \caption{Italy}
  \label{fig:IT_CISS}

\includegraphics[width=\maxwidth]{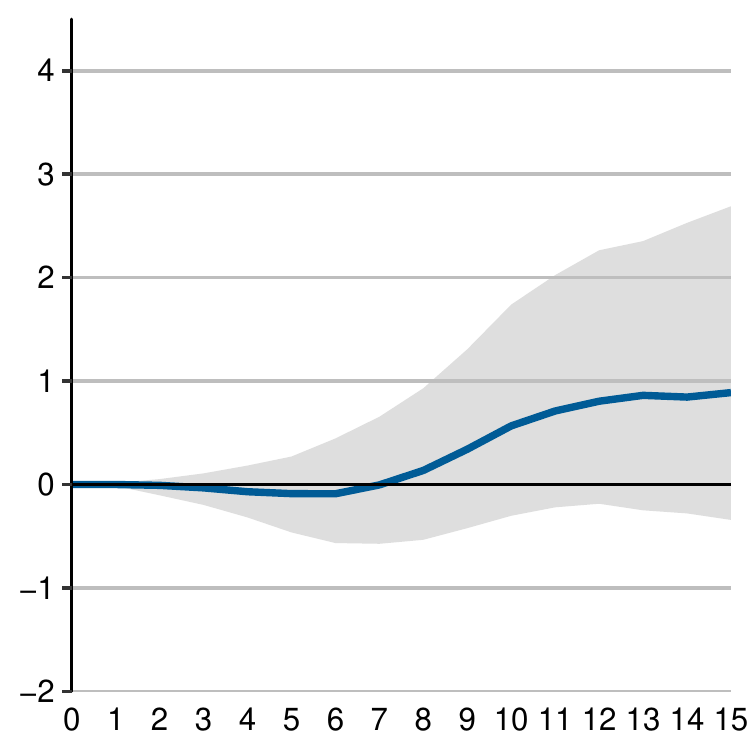} 

  \end{subfigure}
  \begin{subfigure}{.24\textwidth}
  \caption{Portugal}
  \label{fig:PT_CISS}

\includegraphics[width=\maxwidth]{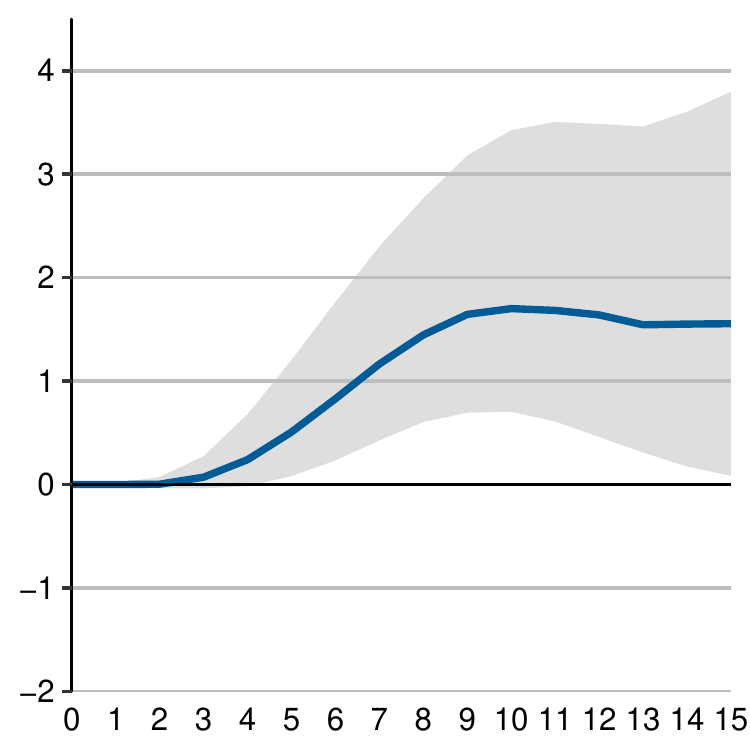} 

  \end{subfigure}
  \begin{subfigure}{.24\textwidth}
  \caption{Spain}
  \label{fig:ES_CISS}

\includegraphics[width=\maxwidth]{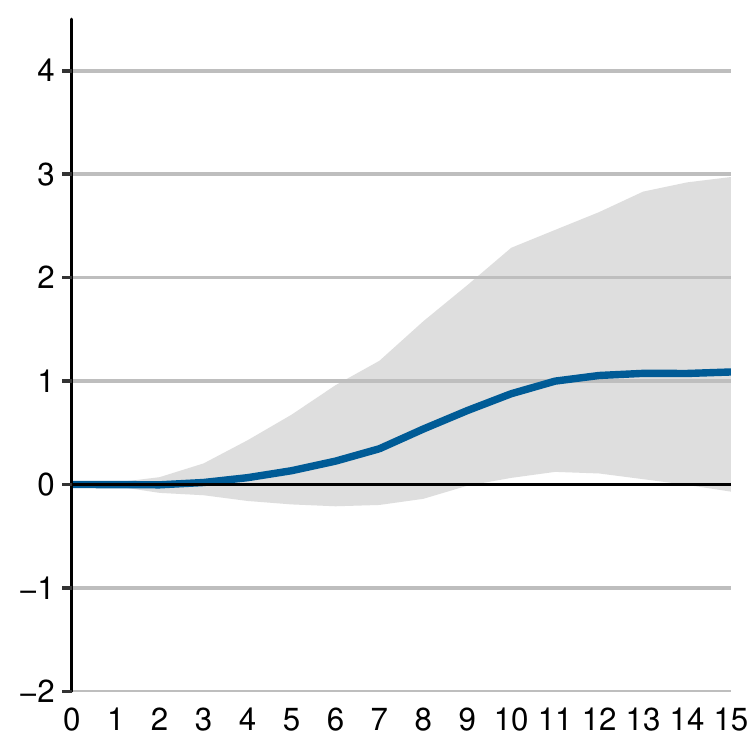} 

  \end{subfigure}
\caption{Impulse Responses of CISS to an EONIA Shock. \textit{Note}: The chart plots the responses of the Composite Indicator of Systemic Stress (CISS) to a expansionary 25 basis point monetary policy shock in the form of a decrease of the EONIA rate. The responses are divided by the shock itself and thus normalized, which yields a percentage point representation on the y-axis. The \textit{solid blue line} represents the median response along the \textit{grey shaded area} which covers the Bayesian credible interval between the 16th and 84th quantile. The x-axis steps are monthly.}
\label{fig:irf_CISS}
\end{figure}

%Gov. Bonds
\begin{figure}[h!]
  \begin{subfigure}{.12\textwidth}

  \caption*{}
  \label{fig:dummy11}
  \end{subfigure}
  \begin{subfigure}{.24\textwidth}
  \caption{Austria}
  \label{fig:AT_GovB}

\includegraphics[width=\maxwidth]{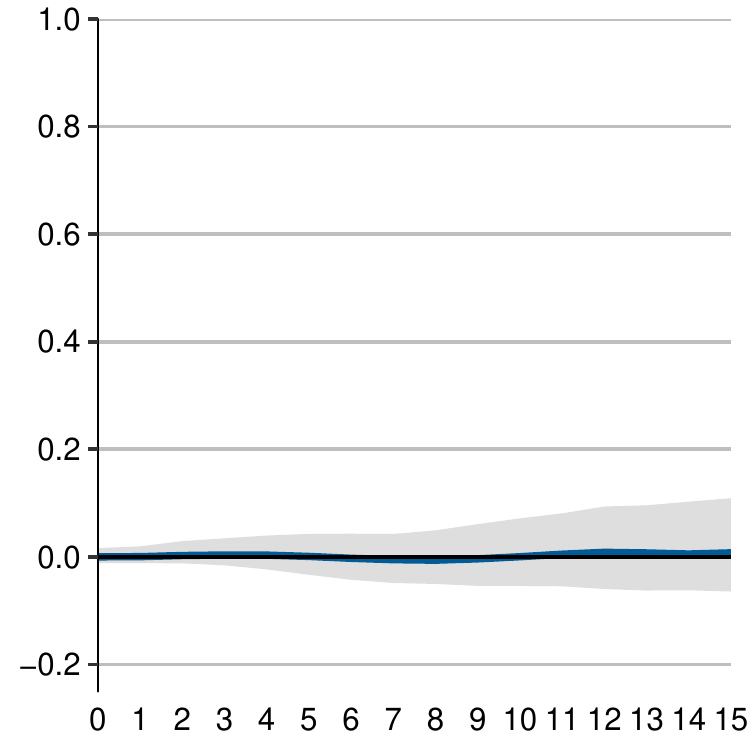} 

  \end{subfigure}
  \begin{subfigure}{.24\textwidth}
  \caption{Belgium}
  \label{fig:BE_GovB}

\includegraphics[width=\maxwidth]{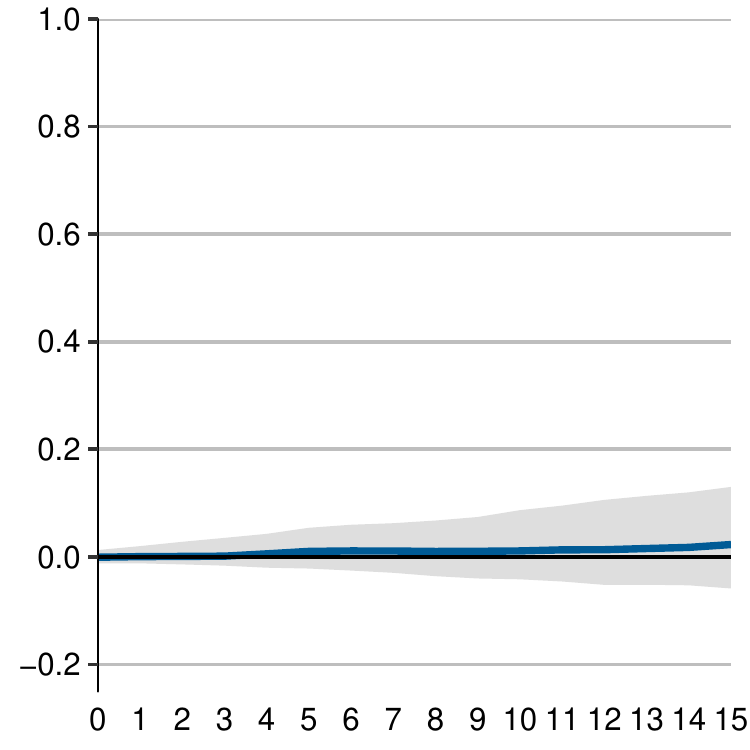} 

  \end{subfigure}
  \begin{subfigure}{.24\textwidth}
  \caption{Germany}
  \label{fig:DE_GovB}

\includegraphics[width=\maxwidth]{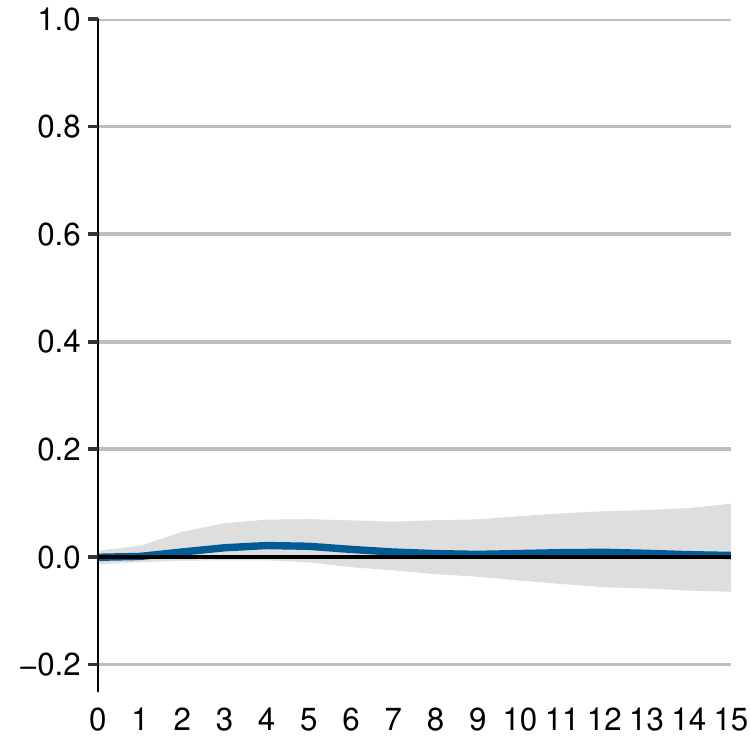} 

  \end{subfigure}
  \begin{subfigure}{.12\textwidth}

  \caption*{}
  \label{fig:dummy12}
  \end{subfigure}
  
  \vspace{0.5cm}
  
  \begin{subfigure}{.24\textwidth}
  \caption{Greece}
  \label{fig:GR_GovB}

\includegraphics[width=\maxwidth]{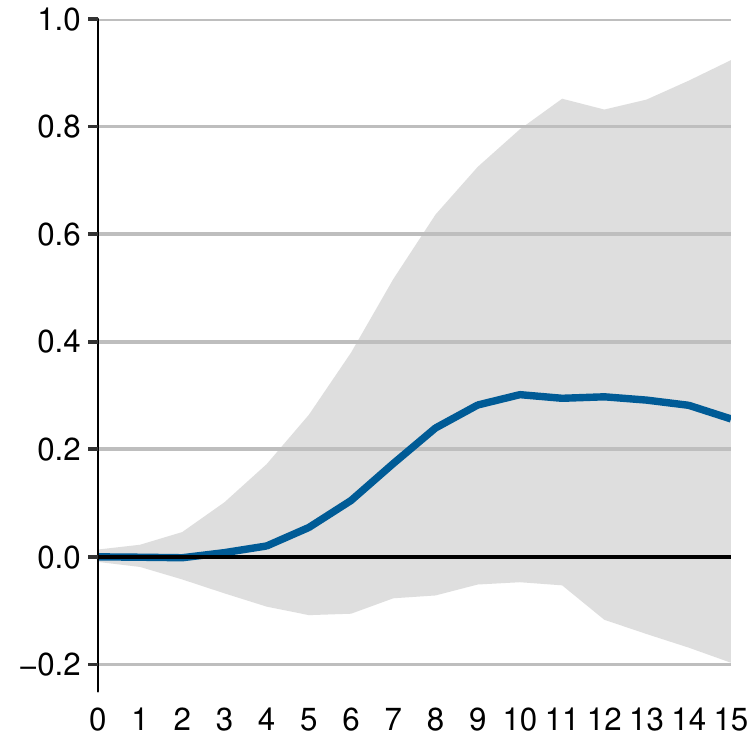} 

  \end{subfigure}
    \begin{subfigure}{.24\textwidth}
  \caption{Italy}
  \label{fig:IT_GovB}

\includegraphics[width=\maxwidth]{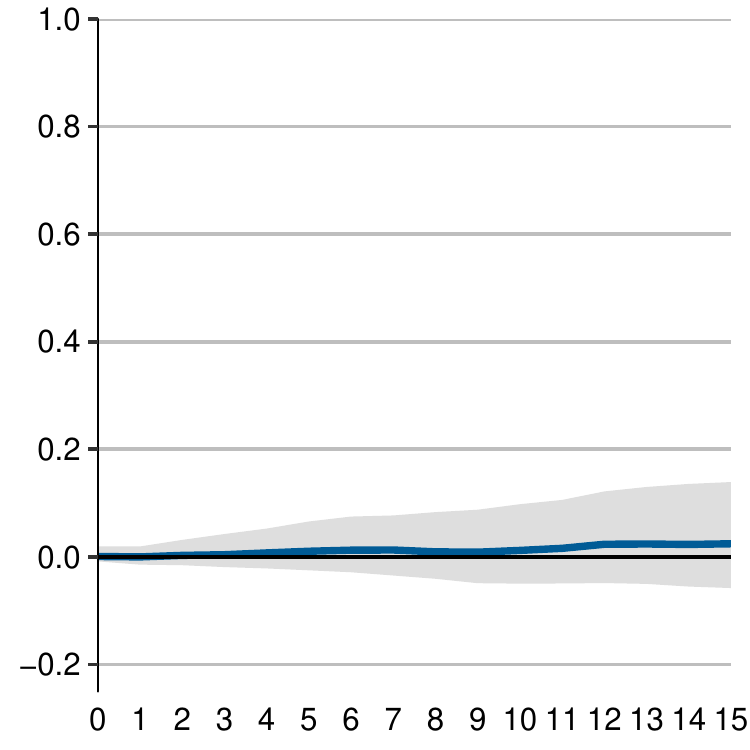} 

  \end{subfigure}
  \begin{subfigure}{.24\textwidth}
  \caption{Portugal}
  \label{fig:PT_GovB}

\includegraphics[width=\maxwidth]{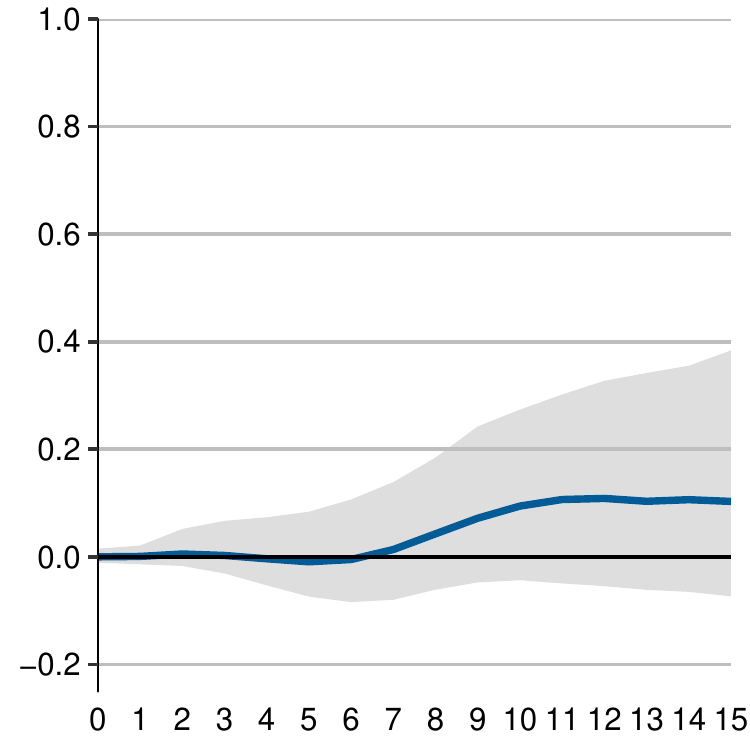} 

  \end{subfigure}
  \begin{subfigure}{.24\textwidth}
  \caption{Spain}
  \label{fig:ES_GovB}

\includegraphics[width=\maxwidth]{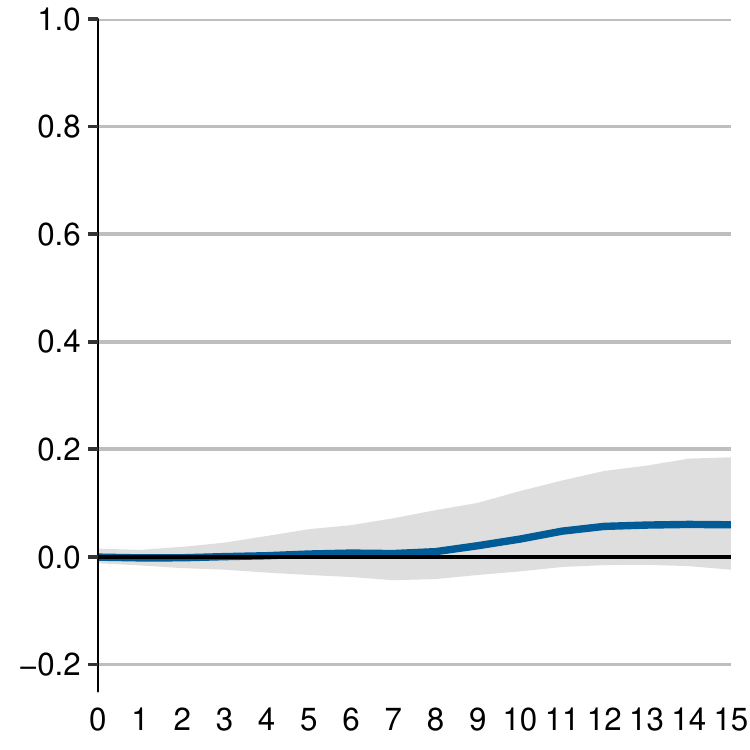} 

  \end{subfigure}
\caption{Impulse Responses of Gov. Bond Yields to an EONIA Shock. \textit{Note}: The chart plots the responses of the long-term government bond yields to a expansionary 25 basis point monetary policy shock in the form of a decrease of the EONIA rate. The responses are divided by the shock itself and thus normalized, which yields a percentage point representation on the y-axis. The \textit{solid blue line} represents the median response along the \textit{grey shaded area} which covers the Bayesian credible interval between the 16th and 84th quantile. The x-axis steps are monthly.}
\label{fig:irf_GovB}
\end{figure}

%Deposits
\begin{figure}[h!]
  \begin{subfigure}{.12\textwidth}

  \caption*{}
  \label{fig:dummy13}
  \end{subfigure}
  \begin{subfigure}{.24\textwidth}
  \caption{Austria}
  \label{fig:AT_Deposits}

\includegraphics[width=\maxwidth]{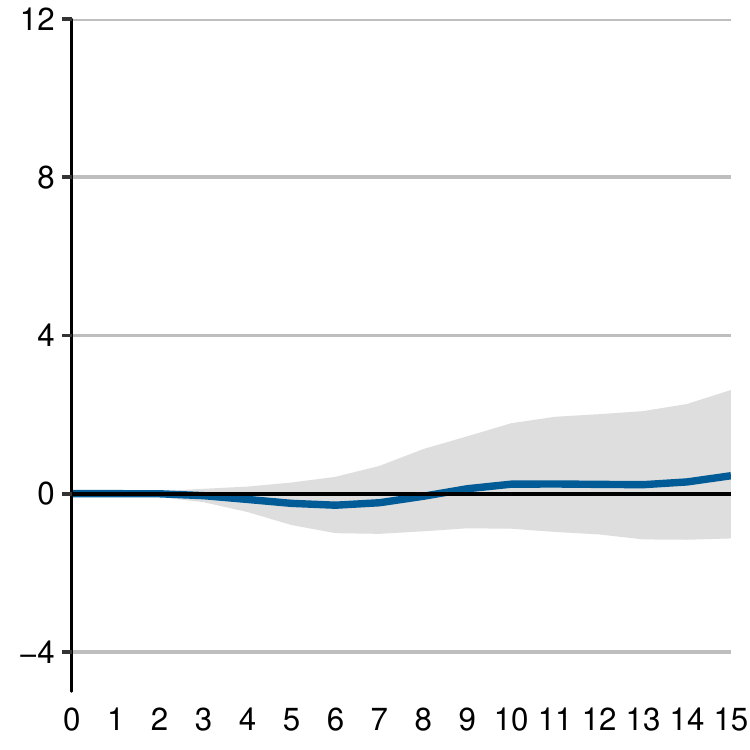} 

  \end{subfigure}
  \begin{subfigure}{.24\textwidth}
  \caption{Belgium}
  \label{fig:BE_Deposits}

\includegraphics[width=\maxwidth]{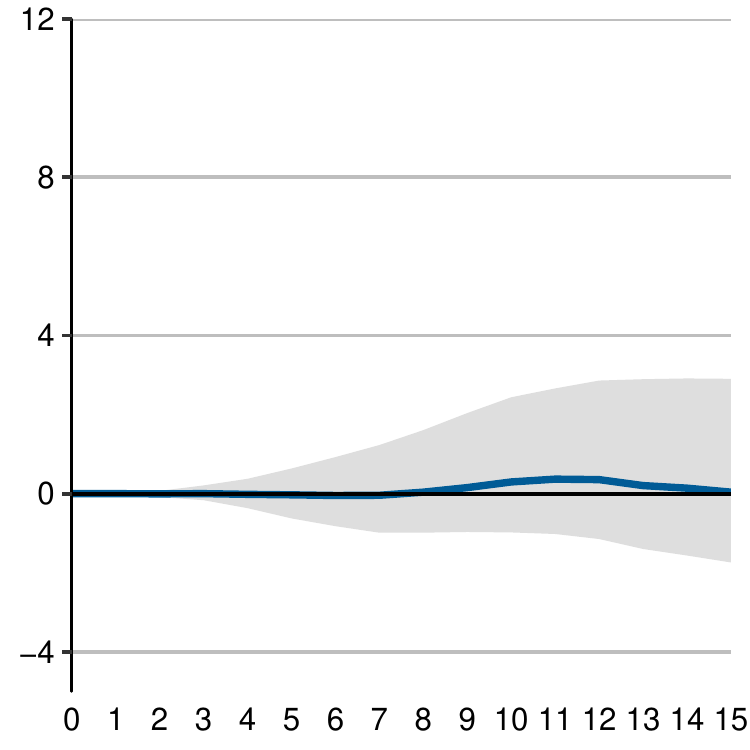} 

  \end{subfigure}
  \begin{subfigure}{.24\textwidth}
  \caption{Germany}
  \label{fig:DE_Deposits}

\includegraphics[width=\maxwidth]{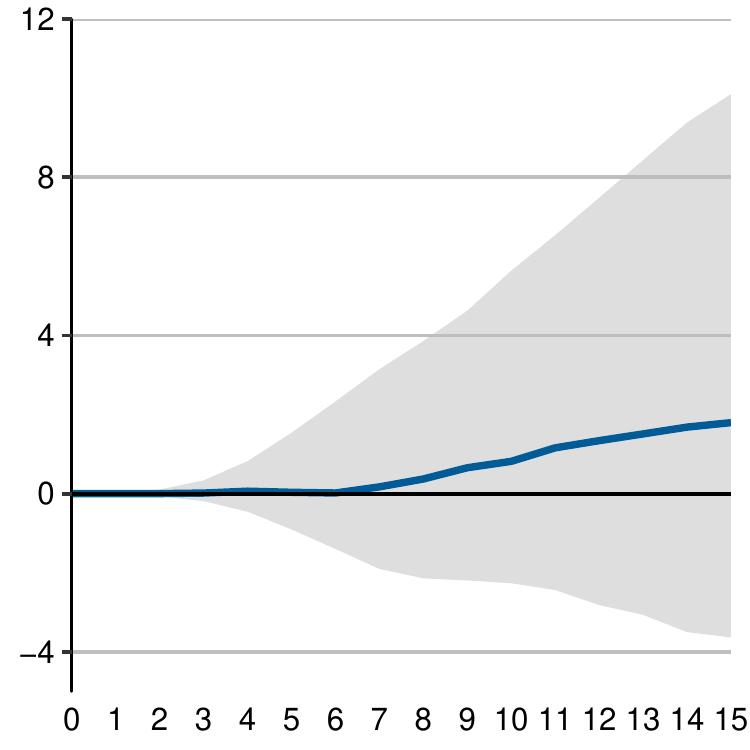} 

  \end{subfigure}
  \begin{subfigure}{.12\textwidth}

  \caption*{}
  \label{fig:dummy14}
  \end{subfigure}
  
  \vspace{0.5cm}
  
  \begin{subfigure}{.24\textwidth}
  \caption{Greece}
  \label{fig:GR_Deposits}

\includegraphics[width=\maxwidth]{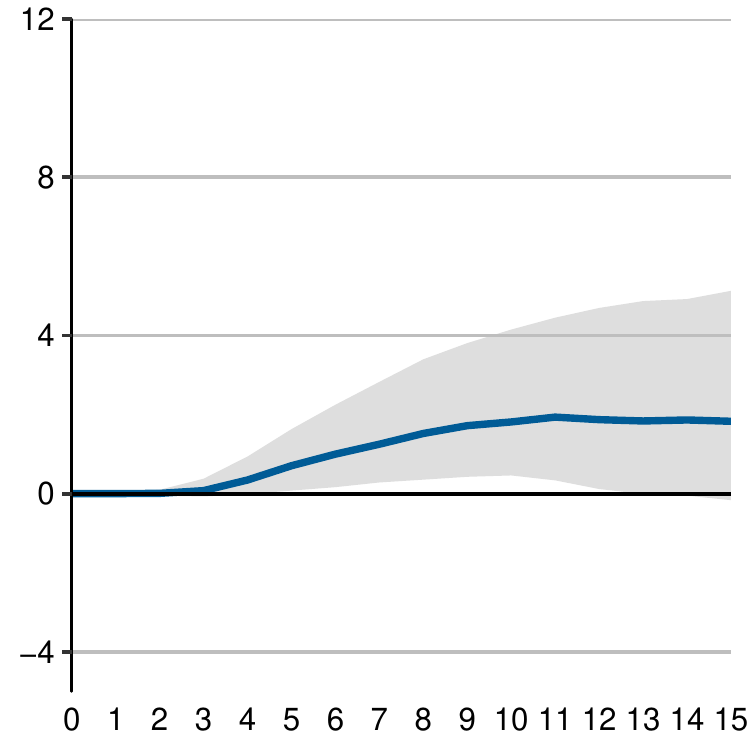} 

  \end{subfigure}
    \begin{subfigure}{.24\textwidth}
  \caption{Italy}
  \label{fig:IT_Deposits}

\includegraphics[width=\maxwidth]{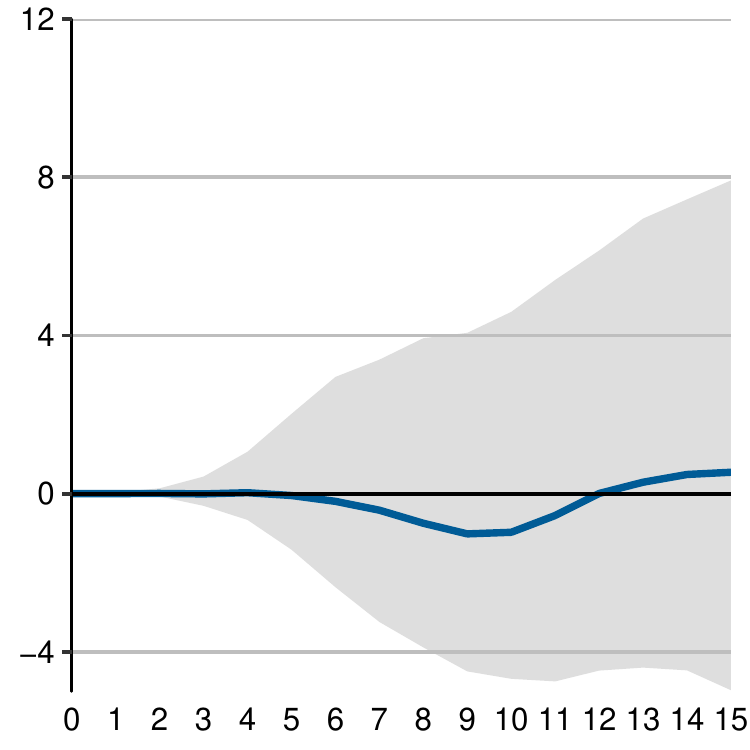} 

  \end{subfigure}
  \begin{subfigure}{.24\textwidth}
  \caption{Portugal}
  \label{fig:PT_Deposits}

\includegraphics[width=\maxwidth]{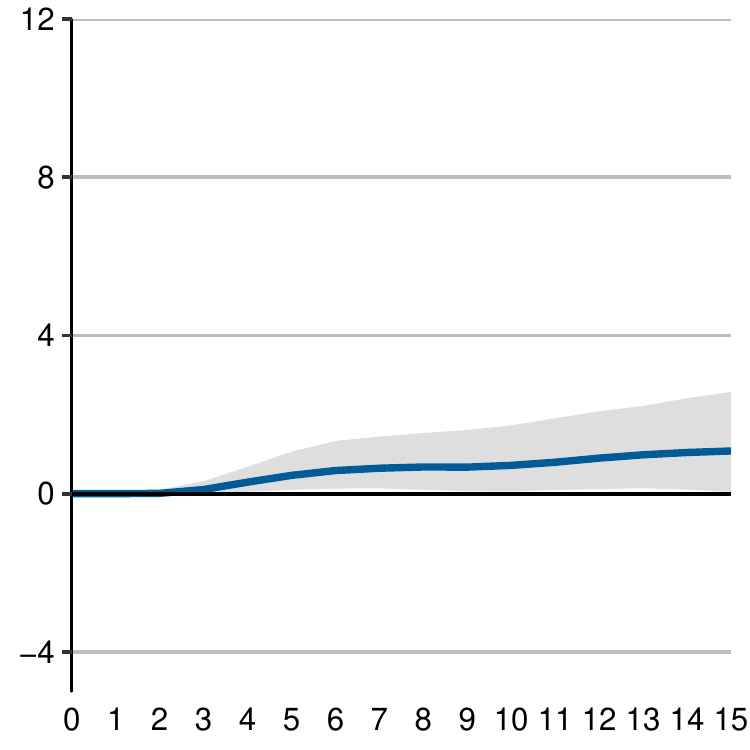} 

  \end{subfigure}
  \begin{subfigure}{.24\textwidth}
  \caption{Spain}
  \label{fig:ES_Deposits}

\includegraphics[width=\maxwidth]{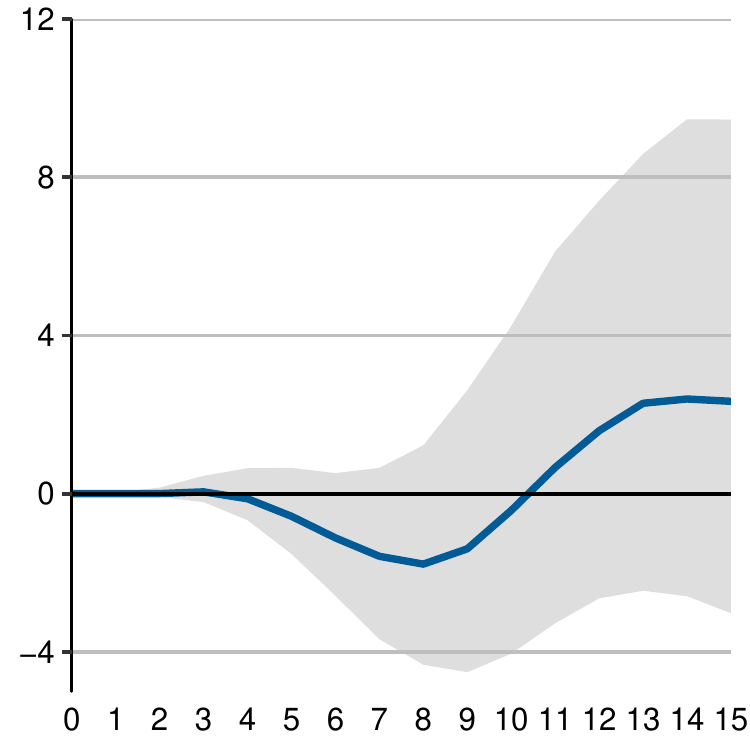} 

  \end{subfigure}
\caption{Impulse Responses of MFI Deposits to an EONIA Shock. \textit{Note}: The chart plots the responses of vis-a-vis deposits of monetary financial institution (MFI) to a expansionary 25 basis point monetary policy shock in the form of a decrease of the EONIA rate. The responses are divided by the shock itself and thus normalized, which yields a percentage point representation on the y-axis. The \textit{solid blue line} represents the median response along the \textit{grey shaded area} which covers the Bayesian credible interval between the 16th and 84th quantile. The x-axis steps are monthly.}
\label{fig:irf_Deposits}
\end{figure}

%EUROAREA PLOTS
\begin{figure}[h!]
  \begin{subfigure}{.12\textwidth}

  \caption*{}
  \label{fig:dummy15}
  \end{subfigure}
  \begin{subfigure}{.36\textwidth}
  \caption{Euro Area - MRO}
  \label{fig:EA_MRO}

\includegraphics[width=\maxwidth]{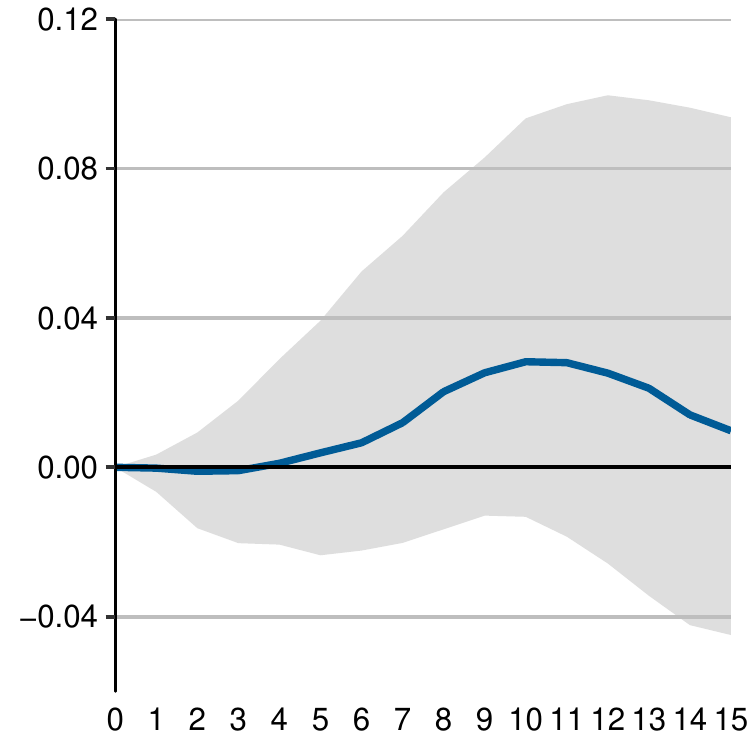} 

  \end{subfigure}
  \begin{subfigure}{.36\textwidth}
  \caption{Euro Area - Stoxx}
  \label{fig:EA_STOXX}

\includegraphics[width=\maxwidth]{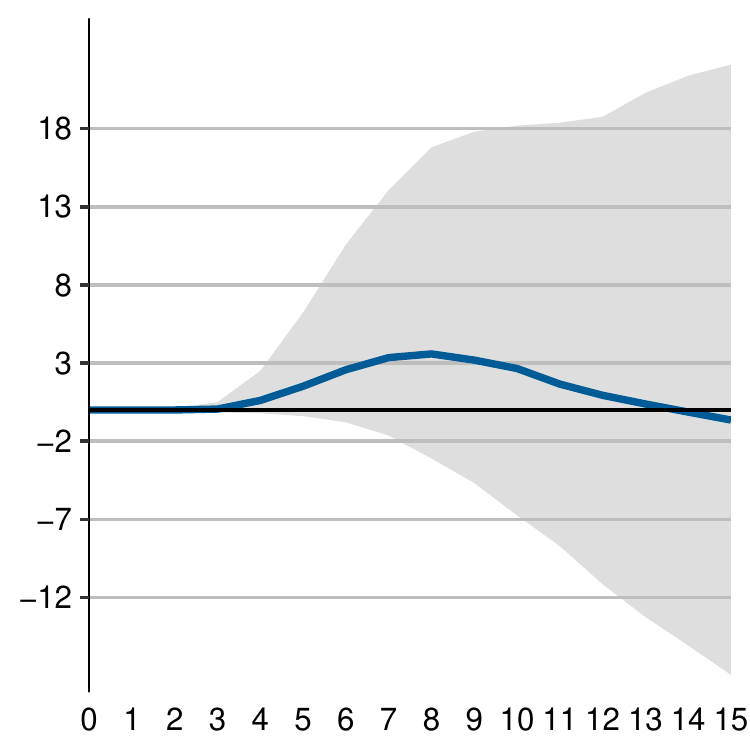} 

  \end{subfigure}
    \begin{subfigure}{.12\textwidth}

  \caption*{}
  \label{fig:dummy16}
  \end{subfigure}
\caption{Impulse Responses of MRO rate and Euro Stoxx to an EONIA Shock. \textit{Note}: The chart plots the responses of the main refinancing operations (MRO) rate and the Euro Stoxx index to a expansionary 25 basis point monetary policy shock in the form of a decrease of the EONIA rate. The responses are divided by the shock itself and thus normalized, which yields a percentage point representation on the y-axis. The \textit{solid blue line} represents the median response along the \textit{grey shaded area} which covers the Bayesian credible interval between the 16th and 84th quantile. The x-axis steps are monthly.}
\label{fig:irf_EA}
\end{figure}

\end{appendices}

\end{document}